\documentclass[11pt,a4paper]{article}
\pdfoutput=1
\usepackage{jheppub}
\usepackage{color}
\usepackage{graphicx}
\usepackage{wrapfig}
\usepackage{verbatim}
\usepackage{amsmath}
\usepackage{amssymb}
\usepackage{url}
\usepackage{bbold}
\usepackage{xspace}
\usepackage{slashed}
\usepackage{multirow}
\usepackage{threeparttable}
\usepackage{paralist}
\usepackage{color}
\usepackage{subcaption}
\usepackage{floatrow}
\newfloatcommand{capbtabbox}{table}[][\FBwidth]

\usepackage{tikz}
\usetikzlibrary{trees}
\usetikzlibrary{decorations.pathmorphing}
\usetikzlibrary{decorations.markings}
\usetikzlibrary{arrows}

\definecolor{darkgreen}{rgb}{0,0.5,0}

\newcommand{\TeV}{\text{TeV}}
\newcommand{\SM}{\text{SM}}

\newcommand{\Lagr}{\mathcal{L}}

\newcommand{\hc}{\text{h.c}}

\DeclareRobustCommand{\Sec}[1]{Sec.~\ref{#1}}
\DeclareRobustCommand{\Secs}[2]{Secs.~\ref{#1} and \ref{#2}}
\DeclareRobustCommand{\App}[1]{App.~\ref{#1}}
\DeclareRobustCommand{\Tab}[1]{Table~\ref{#1}}

\DeclareRobustCommand{\Fig}[1]{Fig.~\ref{#1}}

\DeclareRobustCommand{\Eq}[1]{Eq.~(\ref{#1})}
\DeclareRobustCommand{\Eqs}[2]{Eqs.~(\ref{#1}) and (\ref{#2})}
\DeclareRobustCommand{\Ref}[1]{Ref.~\cite{#1}}

\newcommand{\Dfbd}{\mathord{\buildrel{\lower3pt\hbox{$\scriptscriptstyle\leftrightarrow$}}\over {D}_{\mu}}}

\newcommand{\beq}{\begin{equation}}
\newcommand{\eeq}[1]{\label{#1}\end{equation}}
\def\beqa{\begin{eqnarray}}
\def\eeqa#1{\label{#1}\end{eqnarray}}
\newcommand{\eeqn}{\end{equation}}

\newcommand{\ldm}{\lambda_{\text{\tiny DM}}}
\newcommand{\Ltop}{\Lagr_{\text{\tiny top}}}

\def\stacksymbols #1#2#3#4{\def\theguybelow{#2}
    \def\vp{\lower#3pt}
    \def\sp{\baselineskip0pt\lineskip#4pt}
    \mathrel{\mathpalette\intermediary#1}}

\def\intermediary#1#2{\vp\vbox{\sp
     \everycr={}\tabskip0pt
     \halign{$\mathsurround0pt#1\hfil##\hfil$\crcr#2\crcr
              \theguybelow\crcr}}}

\begin{document}
{\large
\flushright TUM-HEP 1092/17 \\
}
\title{Little composite dark matter} 

\author[a]{Reuven Balkin,}
\author[b]{Gilad Perez,}
\author[a]{Andreas Weiler}

\affiliation[a]{Physik-Department, Technische Universit\"at M\"unchen, 85748 Garching, Germany}
\affiliation[b]{Department of Particle Physics and Astrophysics, Weizmann Institute of Science, Rehovot 7610001, Israel}
\emailAdd{reuven.balkin@tum.de}
\emailAdd{gilad.perez@weizmann.ac.il}
\emailAdd{andreas.weiler@tum.de }

\date{\today}

\abstract{We examine the dark matter phenomenology of a composite electroweak singlet state. This singlet belongs to the Goldstone sector of a well-motivated extension of the Littlest Higgs with $T$-parity. A viable parameter space, consistent with the observed dark matter relic abundance as well as with the various collider, electroweak precision and dark matter direct detection experimental constraints is found for this scenario. $T$-parity implies a rich LHC phenomenology, which forms an interesting interplay between conventional natural SUSY type of signals involving third generation quarks and missing energy, from stop-like particle production and decay, and composite Higgs type of signals involving third generation quarks associated with Higgs and electroweak gauge boson, from vector-like top-partners production and decay. The composite features of the dark matter phenomenology allows the composite singlet the produce the correct relic abundance while interacting weakly with the Higgs via the usual Higgs portal coupling $\ldm \sim O(1\%)$, thus evading direct detection.  
}

%\keywords{}

%\arxivnumber{14xx.xxxx}

\preprint{}
\maketitle
\section{Introduction}
The Hierarchy problem of the Standard Model (SM) could be solved by assuming that the Higgs is a pseudo Nambu-Goldstone Boson (pNGB) of a spontaneously broken global symmetry~\cite{Kaplan:1983fs,Georgi:1984af,Kaplan:1983sm,Dugan:1984hq}. In this scenario, the Higgs is not an elementary particle but rather a composite state whose constituents are held together by some new strong force. In this respect, the Composite Higgs resembles other scalars found in nature, the QCD pions. An extended composite sector could also explain the origin of dark matter (DM)~\cite{Frigerio:2012uc,Marzocca:2014msa,Ballesteros:2017xeg,Ma:2017vzm,Balkin:2017aep}. The same strong dynamics responsible for the Higgs may produce a stable neutral scalar bound state, a composite DM candidate. This could be considered in analogy to the Proton, another QCD bound state, which is an abundant particle in our universe, whose stability is insured by an (accidental) global symmetry. The composite DM candidate is a pNGB and it could be naturally as light as the weak scale, which fits in the weakly interacting massive particle (WIMP) paradigm. 
\medskip 

\noindent
One realization of the composite Higgs scenario is the Littlest Higgs~\cite{ArkaniHamed:2002qy,ArkaniHamed:2002qx,Low:2002ws,Chang:2003un,Skiba:2003yf,Chang:2003zn,Schmaltz:2010ac}. The original model is strongly constrained by electroweak precision tests (EWPT) due to tree level contributions to electroweak observables~\cite{Csaki:2002qg,Csaki:2003si,Hewett:2002px,Han:2003wu,Gregoire:2003kr,Casalbuoni:2003ft,Kilic:2003mq,Kilian:2003xt}. These constraints required the symmetry breaking scale $f$ to be a few TeV, thus reintroducing considerable fine-tuning.  $T$-Parity has been proposed in order to prevent tree-level exchanges of heavy states~\cite{Cheng:2003ju,Cheng:2004yc,Low:2004xc,Hubisz:2005tx}. The new heavy states are odd under a discrete $T$-parity,  therefore contributions to electroweak observables are possible only at the 1-loop level. This allows the symmetry breaking scale $f$ to be $O(1)$ TeV. As an added benefit, $T$-Parity can be used as a stabilizing symmetry for a DM candidate, as the lightest $T$-odd particle is guaranteed to be stable.  
\medskip 
 
\noindent
In this work, we consider the phenomenology of a Littlest Higgs model with $T$-parity (LHT) with a consistent implementation of $T$-parity in the fermionic sector~\cite{Pappadopulo:2010jx}.
Compared to the simplest LHT model, one enlarges the symmetry breaking pattern and also the unbroken symmetry group $H$ which allows a complete composite representation containing just one fermion doublet.
 In particular, we analyze the DM phenomenology of a composite singlet scalar. In \Sec{model} we present the model and motivate the extension leading to the larger Goldstone sector. In \Sec{ScalarPotentialSec} we briefly review the scalar potential structure. A detailed discussion of the scalar potential can be found in \App{scalarPotential}. In \Secs{lhcPhen}{ewpt} we derive constraints on the model parameters from recent LHC searches and EWPT. In \Sec{dmPheno} we discuss the DM phenomenology of the composite singlet DM. We finally summarize our results and conclude  in \Sec{conc}.

\section{Model} \label{model}
The model is based on the Littlest Higgs (LH) $SU(5)/SO(5)$ non-linear sigma model~\cite{ArkaniHamed:2002qy}. We define the scalar field $\Sigma$ in the symmetric $\mathbf{15}$ representation of $SU(5)$:
\begin{align}
\Sigma \to U \Sigma U^T\,.
\end{align}
$\Sigma$ develops a vacuum expectation value (VEV), 
\begin{align}
\left< \Sigma \right> \equiv \Sigma_0 = \begin{pmatrix}
& & \mathbb{1}_2 \\
& 1 & \\
\mathbb{1}_2 & &
\end{pmatrix}\,,
\end{align}
spontaneously breaking $SU(5)$ to $SO(5)$. The 10 unbroken SO(5) generators denoted by $T_i$ satisfy
\begin{align}
T_i\Sigma_0  + \Sigma_0 T^T_i = 0\,,
\end{align}
and the 14 broken SU(5) generators denoted by $X_j$  satisfy
\begin{align}
X_j\Sigma_0  - \Sigma_0 X^T_j = 0\,.
\end{align}
We gauge two subgroups of $SU(5)$, denoted by $[SU(2)\times U(1)]_i$ with $i=1,2$. The gauged generators are
\begin{align}
Q_1^a \equiv \begin{pmatrix}
 \sigma^a/2 & 0 & 0\\
0& 0& 0 \\
0& 0& 0 \\
\end{pmatrix}\,, \;\;\; Y_1 =\text{Diag}(3,3,-2,-2,-2)/10\,,  \label{su2l}
\\
Q_2^a \equiv \begin{pmatrix}
0 & 0 & 0\\
0& 0& 0 \\
0& 0& -{\sigma^a}^*/2  \\
\end{pmatrix}\,, \;\;\; Y_2 =\text{Diag}(2,2,2,-3,-3)/10\,. \label{su2r}
\end{align}
$\sigma^a$ with $a=1,2,3$ are the Pauli matrices. $\Sigma_0$ spontaneously breaks the gauge symmetry to its diagonal subgroup $[SU(2)\times U(1)]_{1+2}$, which we identify as the SM electroweak gauge group.   The Nambu-Goldstone bosons (NGB's) associated with the broken $SU(5)$ generators decompose under the SM gauge group to the following representations
\begin{align}
\bf{1}_0 \oplus \bf{3}_0 \oplus \bf{2}_{1/2} \oplus \bf{3}_{1}\,. \label{originalCoset}
\end{align}
We can parameterize the low energy degrees of freedom of the $\Sigma$ field using the NGB's, defining $\Pi_\Sigma \equiv \pi_a X_a$:
\begin{align}
\Sigma &= e^{i\Pi_\Sigma /f} \Sigma_0 e^{i\Pi^T_\Sigma /f} = e^{2i\Pi_\Sigma /f} \Sigma_0\,, \nonumber
\\ \Pi_\Sigma &= \begin{pmatrix}
\frac{\tau \cdot \sigma}{2}+ \frac{\phi_0}{2 \sqrt{5}}  \mathbb{1}_2  & \frac{H}{\sqrt{2}} & \Phi \\
 \frac{H^\dagger}{\sqrt{2}} &- \frac{2\phi_0}{ \sqrt{5}} &  \frac{H^T}{\sqrt{2}} \\
\Phi^\dagger &  \frac{H^*}{\sqrt{2}} & \frac{\tau \cdot \sigma^*}{2}+ \frac{\phi_0}{2 \sqrt{5}}  \mathbb{1}_2
\end{pmatrix}\,,\;\; \text{with}\,\,\, \Phi = \begin{pmatrix}
\Phi^{++} & \Phi^+ / \sqrt{2} \\
\Phi^+ / \sqrt{2}  & \Phi^0
\end{pmatrix} \label{pi_sigma matrix}\,.
\end{align}
In the original LH model, the triplet $\tau$ and the singlet $\phi_0$ are "eaten" by the heavy gauge bosons. The physical scalar spectrum contains the complex doublet $H$ which we identify as the SM Higgs field, and a heavy charged triplet $\Phi$. The gauge spectrum contains the SM gauge fields and additional heavy gauge fields with masses $m_{W_H}\sim gf\,, m_{B_H}\sim g' f\,,$ with $g,g'$ the SM gauge couplings.  The heavy gauge states contribute at tree level to the electroweak oblique parameters. These contributions lead to stringent constraints from electroweak precision tests (EWPT), pushing the symmetry breaking scale of the original LH model $f \sim \text{a few TeV}$ (e.g \Ref{Csaki:2002qg}). The corrections to electroweak observables from the heavy gauge states are made smaller by introducing a discrete symmetry which forbids tree level exchanges of heavy states. The addition of a discrete symmetry stabilizes the lightest odd particle, making it a viable DM candidate. 
This discrete symmetry, usually referred to as $T$-parity,  is defined as~\cite{Cheng:2003ju}
\begin{align}
\text{$T$-parity: }\,\, T_i \to \Omega  T_i \Omega \,, \;\;\; X_j \to -\Omega  X_j\Omega
\end{align}
with
\begin{align}
\Omega  = -\exp[2 \pi i Q^3_{1+2} ] = \text{diag}(1,1,-1,1,1)\,, \label{Tparitydef}
\end{align}
which is an automorphism defined on the $SU(5)$ generators. This definition determines the $T$-parity of all the fields associated with the $SU(5)$ generators, namely the Goldstone and gauge fields. The $\Omega$ rotation is introduced to make the Higgs even under $T$-parity, while keeping the rest of the Goldstone fields odd. For the gauge fields, the $T$-parity transformation can be interpreted as an exchange symmetry between the gauge groups $1 \leftrightarrow 2$. Hence the diagonal combination is even, and the broken combination is odd. 
\\ \\ Let us understand how linear representations of $SU(5)$ transform under $T$-parity. One can use \Eq{Tparitydef} to show that each transformation $g = e^{i \alpha_j X_j + i \beta_i T_i}\in SU(5)$ is mapped under $T$-parity to
\begin{align}
g ~\to~ \tilde{g} \equiv \Omega  \Sigma_0 g^* \Sigma_0 \Omega \,.
\end{align}
Therefore, up to a constant matrix, fundamental and anti-fundamental indices of $SU(5)$ are mapped to each other 
\begin{align}
\underbrace{V_i}_{\bf{5}} ~\leftrightarrow ~   (\Sigma_0 \Omega)_{ij}\underbrace{U^j}_{\overline{\bf{5}}}\,.
\end{align}
The $\Sigma$ field transforms with two fundamental $SU(5)$ indices, so under $T$-parity
\begin{align}
\Sigma ~\to ~\tilde\Sigma \equiv  \Omega  \Sigma_0  \Sigma^\dagger \Sigma_0 \Omega  \,. \label{sigmatilde}
\end{align}
\subsection{A UV doubling problem, making the $T$-odd doublet massive}
The coset structure of LH with $T$-parity is in tension with the SM matter content~\cite{Csaki:2008se,Pappadopulo:2010jx}. The low energy theory must contain a $T$-even massless $SU(2)$ doublet, the left-handed quark doublet of the SM. Since $T$-parity can be understood as an exchange symmetry between the two gauged $SU(2)$ subgroups of $SU(5)$ (we omit the $U(1)$ factors for the following discussion), one must therefore introduce two doublets $\psi_i$, each transforming under a different $SU(2)_i$ with $i=1,2$. Under $T$-parity the two doublets are mapped into each other
\begin{align}
\psi_1 ~\leftrightarrow~  \psi_2\,.
\end{align}
We would like to write a mass term for the $T$-odd combination $\psi_- \equiv (\psi_1-\psi_2)$ that respects the SM gauge group. Let us introduce a right-handed field $\psi^c$ transforming as a doublet under the SM gauge group $[SU(2)]_{1+2}$
\begin{align}
\Lagr \ni (\overline{\psi}_1-\overline{\psi}_2)\psi^c\,. \label{oddMassTerm}
\end{align}
This term respects the SM gauge group, however each term by itself breaks $SU(2)_1 \times SU(2)_2$ and cannot be generated by a reasonable UV theory which respects those gauge symmetries, unless they are spontaneously broken. Assuming that $\psi^c$ cannot be a doublet of just one of the $SU(2)'s$,  we expect the mass term to arise as a result of spontaneous symmetry breaking 
\begin{align}
\Lagr \ni (\overline{\psi}_1\left< \phi_1\right>-\overline{\psi}_2\left< \phi_2\right>)\psi^c\,, \label{toymassterm}
\end{align}
where we introduced two sources of spontaneous symmetry breaking, the VEV's $\left< \phi_1\right>$ and $\left< \phi_2\right>$. 

\noindent
Let us examine now the VEV's which we can use to write this term in a gauge-invariant way. In \Secs{sigma0}{thirdGaugeGroup} we briefly examine two different constructions presented in the literature that generate the mass term of \Eq{oddMassTerm}.  We mention possible shortcomings of these constructions, which motivate the construction used in this work, presented in \Sec{mirror}. Readers interested only in the details of the model used in this work, may skip directly to \Sec{mirror}.
\subsubsection{Non-linear formulation of a massive odd doublet} \label{sigma0}
One construction commonly presented in the literature uses the CCWZ formalism~\cite{Coleman:1969sm,Callan:1969sn}. The main advantage of this approach is that no new sources of spontaneous symmetry breaking are needed. First we have the linear representations of $SU(5)$~\cite{Cheng:2004yc}
\begin{align}
\Psi_1 = \begin{pmatrix}
\psi_1 \\ 0 \\0
\end{pmatrix}_{\overline{\bf{5}}}\,,
\;\;
\Psi_2 = \begin{pmatrix}
0 \\ 0 \\ \psi_2
\end{pmatrix}_{\bf{5}}\,, \label{psi1psi2}
\end{align}
with the following $T$-parity transformation 
\begin{align}
\Psi_1 ~\to~ \Omega \Sigma_0 \Psi_2\,.
\end{align}
A mass term for the $T$-odd combination is constructed using a non-linearly transforming field
\begin{align}
\tilde\Psi^c = \begin{pmatrix}
\psi^c_1 \\
\chi^c \\
\psi^c_2
\end{pmatrix}\,,\;\; \text{ a }{\bf{5}} \text{ of } SO(5)\,.
\end{align}
Under a transformation $g\in SU(5)$
\begin{align}
\tilde\Psi^c ~\to~ O(\Pi_\Sigma,g) \tilde\Psi^c\,, \;\;O \in SO(5)\,.
\end{align}
$e^{i\Pi_\Sigma/f}$ transforms under a transformation $g\in SU(5)$ in the following way
\begin{align}
e^{i\Pi_\Sigma/f} ~\to~ g e^{i\Pi_\Sigma/f} O^\dagger  = O e^{i\Pi_\Sigma/f} (\Sigma_0 g^T \Sigma_0)\,.
\end{align}
The kinetic term for $\tilde\Psi^c$ contains the $e_\mu$ symbol defined by~\cite{Coleman:1969sm,Callan:1969sn}
\begin{align}
ie^{-i\Pi_\Sigma/f} (\partial_\mu e^{i\Pi_\Sigma/f}) \equiv d^j_\mu X^{j}+e^i_\mu T^{i}\,.
\end{align}
Using the automorphism defined in \Eq{Tparitydef} we can write $e_\mu \equiv e^i_\mu T^{i}$ in a $T$-parity symmetric form
\begin{align}
e_\mu = \frac{i}{2} \left( e^{-i\Pi_\Sigma/f} (\partial_\mu e^{i\Pi_\Sigma/f})  +  e^{i\Pi_\Sigma/f} (\partial_\mu e^{-i\Pi_\Sigma/f})\right) \,.
\end{align}
The $e_\mu$ symbol transform as a covariant derivative
\begin{align}
(\partial_\mu+e_\mu)~\to~ O(\partial_\mu+e_\mu)O^\dagger\,,
\end{align}
which allows us to write an invariant kinetic term for $\tilde\Psi^c $. Note that under $T$-parity 
\begin{align}
e_\mu \to \Omega e_\mu \Omega\,,
\end{align}
therefore the transformation of $\tilde\Psi^c$ under $T$-parity is
\begin{align}
\tilde\Psi^c \to -\Omega\tilde\Psi^c \,.
\end{align}
The benefit of the CCWZ formalism is that the pion matrix can be used to "dress" the field $\tilde\Psi^c$ as linear representations of $SU(5)$, e.g ${\bf{5}}$ and ${\bf{\bar{5}}}$
\begin{align}
e^{i\Pi_\Sigma/f}\tilde\Psi^c ~\to~ g(e^{i\Pi_\Sigma/f}\tilde\Psi^c)\,, 
\end{align}
and
\begin{align}
\,\,\Sigma_0e^{-i\Pi_\Sigma/f}\tilde\Psi^c ~\to~ g^*(\Sigma_0e^{-i\Pi_\Sigma/f}\tilde\Psi^c)\,,
\end{align}
with $g\in SU(5)$.
Finally the mass term is given by~\cite{Cheng:2004yc}
\begin{align}
\Lagr \ni \frac{\kappa f}{\sqrt{2}} (\overline{\Psi}_1  \Sigma_0 e^{-i\Pi_\Sigma/f}-\overline{\Psi}_2  e^{i\Pi_\Sigma/f})\tilde\Psi^c+\hc =  \frac{\kappa f}{\sqrt{2}} \left( \overline\psi_1-\overline\psi_2\right)\psi_2^c+...\,.  \label{oldmassterm}
\end{align}
The field $\tilde\Psi^c$ must be a complete $SO(5)$ representation, otherwise the kinetic term for $\tilde\Psi^c$ would explicitly break the global symmetry protecting the Higgs mass~\cite{Cheng:2004yc}. The field $\psi_1^c$ is still massless at this point. One could formally introduce an additional doublet $\eta$ and write a mass term
\begin{align}
\Lagr \ni M (\bar\eta \psi_1^c + \hc)\,.
\end{align}
This term breaks the global symmetries protecting the Higgs mass, generating $O(M^2)$ contributions to the Higgs mass.   
\subsubsection{Adding a third $SU(2)\times U(1)$} \label{thirdGaugeGroup}
We conclude that the model requires additional structure in order to give mass to the $T$-odd combination without explicit breaking of the global symmetry. One possible solution is to add an additional gauge group~\cite{Csaki:2008se,Brown:2010ke}, denoted by $[SU(2)\times U(1)]_3$. Now $\psi^c$ of \Eq{toymassterm} transforms as a doublet under $[SU(2)\times U(1)]_3$ and the scalars $\phi_i$  transform as a bi-fundamentals of $[SU(2)\times U(1)]_i \times [SU(2)\times U(1)]_3$ with $i=1,2$. This solution introduces new heavy $T$-even gauge fields. The new $T$-even gauge fields can be made heavy by making the coupling constant of the third $SU(2)\times U(1)$ gauge group large, effectively decoupling them from the theory without spoiling the naturalness of the model. One has the choice of how to enlarge the global symmetry to incorporate this additional gauge group. The most naive extension is
\begin{align}
SU(5) \to SU(5) \times [SU(2) \times U(1)]_3
\end{align}
We introduce additional scalars $\Phi_1$ and $\Phi_2$ transform under the enlarged group as $({\bf{\bar{5}}},{\bf{\bar{2}}})$ and $({\bf{5}},{\bf{\bar{2}}})$ respectively (disregarding the $U(1)$ charges), namely
\begin{align}
\Phi_1 \to g^* \Phi_1 g_3^\dagger\,, \;\; \Phi_2 \to g \Phi_2 g_3^\dagger\,,\;\; g\in SU(5)\,,\;\; g_3 \in [SU(2)\times U(1)]_3\,.
\end{align}
Under $T$-parity
\begin{align}
\Phi_1 \to \Sigma_0 \Omega\Phi_2\,, \;\; \psi^c \to -\psi^c\,.
\end{align}
The $T$-odd doublet gets a mass 
\begin{align}
\Lagr \ni \frac{\kappa }{\sqrt{2}} \left( \overline{\Psi}_1 \left<\Phi_1\right> -  \overline{\Psi}_2 \left<\Phi_2 \right> \right) \psi^c  = \frac{\kappa f }{\sqrt{2}} \left( \overline{\psi}_1  -  \overline{\psi}_2 \right) \psi^c+... \,,
\end{align}
after $\Phi_1$ and $\Phi_2$ acquire  VEV's given by 
\begin{align}
\left< \Phi_1\right> = f\begin{pmatrix}
 \mathbb{1}_{2\times2} \\
 0_{1\times2}\\  
 0_{2\times2}\\ 
\end{pmatrix} = \Sigma_0 \Omega\left< \Phi_2\right>\,.
\end{align}
The appearance of $\Phi_1,\Phi_2$ results in a deviation from the original coset structure of the LH, with the altered coset structure
\begin{align}
\frac{SU(5) \times SU(2) \times U(1)}{[SU(2) \times U(1)]_{1+2+3}}\,.
\end{align}
We now identify $[SU(2) \times U(1)]_{1+2+3}$ as the SM gauge group. This coset contains in the original 14 NGB's of the LH coset, and additional 10 NGB's from the spontaneously broken $SO(5)$. These 10 additional states decompose under the SM gauge group as
\begin{align}
\bf{1}_0 \oplus \bf{3}_0 \oplus \bf{2}_{1/2} \oplus \bf{1}_{1/2} \,.
\end{align}
The additional neutral singlet $\bf{1}_0 $ and triplet $ \bf{3}_0$ are "eaten" by the additional $T$-even gauge fields. This naive approach unavoidably introduces additional physical NGB's in the form of a $T$-odd doublet $\bf{2}_{1/2}$ and a $T$-even complex scalar $\bf{1}_{1/2}$ . These states must be made massive without spoiling the symmetry protection of the SM Higgs. Additional NGB's are a generic result of the enlarged global symmetry structure, even more so when the additional $SU(2)$ is a gauged subgroup of a larger global symmetry~\cite{Csaki:2008se}.
\subsubsection{Mirroring the $1 \leftrightarrow 2$ exchange symmetry} \label{mirror}
In this work we consider a concrete solution suggested in \Ref{Pappadopulo:2010jx}.
We extend the global symmetry 
\begin{align}
SU(5) \to SU(5) \times [SU(2)\times U(1)]_L \times [SU(2)\times U(1)]_R\,.
\end{align}
We introduce a scalar field, $X$, which transforms linearly under $[SU(2)\times~U(1)]_L\times [SU(2)\times~U(1)]_R$
\begin{align}
X \to g_L X g_R^\dagger\,.
\end{align}
When the $\Sigma$ and $X$ acquire VEV's, $\left< \Sigma \right> = \Sigma_0$ and $\left<X\right> = \mathbb{1}_2$, the symmetry is spontaneously broken to
\begin{align}
\frac{SU(5)}{SO(5)} \times \frac{[SU(2)\times U(1)]_L \times [SU(2)\times U(1)]_R}{[SU(2)\times U(1)]_V}\,.
\end{align}
We gauge two $SU(2)\times U(1)$ subgroups defined as the combinations $[SU(2)\times U(1)]_{1+L}$ and $[SU(2)\times U(1)]_{2+R}$.  The residual gauge symmetry $[SU(2)\times U(1)]_{1+2+L+R}$ is identified as the SM gauge group. We can parametrise $X$ using the non-linearly transforming Goldstone fields associated with this symmetry breaking,
\begin{align}
X \equiv e^{\frac{i}{f'}\Pi_X}\left<X\right>e^{\frac{i}{f'}\Pi_X} = e^{\frac{2i}{f'}\Pi_X}\,, \;\;\; \Pi_X  = \frac12 \left( \pi_i \sigma^i + \pi_0 \mathbb{1}_2 \right)\,. \label{pi_X matrix}
\end{align}
Note that the symmetry breaking scale $f'$ may be different than $f$, the symmetry breaking scale of the original coset defined in \Eq{pi_sigma matrix}. $T$-parity in the additional coset is realized as an $L~\leftrightarrow~R$ exchange, mirroring the $1 \leftrightarrow 2$ exchange symmetry of the original coset. Under $T$-parity,
\begin{align}
\Pi_X \to -\Pi_X\,,
\end{align}
We  introduce a non-linear representation of $[SU(2)\times U(1)]_L\times [SU(2)\times U(1)]_R$. $\psi^c$ is a doublet of the unbroken subgroup $[SU(2)\times U(1)]_{L+R}$, transforming non-linearly under  $g_L,g_R \in [SU(2)\times U(1)]_L\times [SU(2)\times U(1)]_R$
\begin{align}
\psi^c \to V(\Pi_X,g_L,g_R)\psi^c\,, \;\; V \in [SU(2)\times U(1)]_{L+R}\,.
\end{align}
The transformation properties under $[SU(2)\times U(1)]_L\times [SU(2)\times U(1)]_R$ of $e^{i \Pi_X/f'}$ in this case are
\begin{align}
e^{i \Pi_X/f'} \to g_L e^{i \Pi_X/f'} V^\dagger = V e^{i \Pi_X/f'} g_R^\dagger\,.
\end{align}
This object can be used to "dress" $\psi^c$ as linear representations
\begin{align}
e^{i \Pi_X/f'}\psi^c \to g_L (e^{i \Pi_X/f'}\psi^c)\,, \;\; e^{-i \Pi_X/f'}\psi^c \to g_R (e^{i \Pi_X/f'}\psi^c)\,.
\end{align}
Finally the mass term can be written as~\cite{Pappadopulo:2010jx}
\begin{align}
\Lagr \ni (\overline{\psi}_1 e^{i\Pi_X/f'}-\overline{\psi}_2  \Sigma_0 e^{-i\Pi_X/f'})\psi^c+\hc\,. \label{quasi_kappa}
\end{align}
This extension allows us to add a single $SU(2)$ doublet to the spectrum, $\psi^c$, and write a mass term for the $T$-odd doublet, without any explicit breaking of the global symmetry. In additional to the 14 original NGB's of \Eq{originalCoset}, our spectrum includes now an additional NGB's, a real singlet $\bf{1}_0 $ and a real triplet $\bf{3}_0 $.
\subsection{Gauge sector} 
We write the Lagrangian for the non-linear $\sigma$ model
\begin{align}
\Lagr_{nl\sigma } = \frac{f^2}{8}\text{Tr}[(D_\mu \Sigma)(D^\mu \Sigma^*)]+ \frac{f'}{4}\text{Tr}[(D_\mu X )(D^\mu X^\dagger ) ] \label{nlsigma}
\end{align}
We parameterize $\Sigma,X$ using the NGB's as defined in \Eq{pi_sigma matrix} and \Eq{pi_X matrix}. The exact form of the covariant derivatives can be found in \App{modelApp}
\\Once we set $\Sigma,X$ to their respected VEV's, we find that the following linear combinations,
\begin{align}
W^a_H = \frac{1}{\sqrt{2}}(W_1^a-W_2^a)\,,\;\; B_H = \frac{1}{\sqrt{2}}(B_1-B_2) \,,
\end{align}
acquire a mass
\begin{align}
M^2_{W_H} = g^2 f^2 (1+r^2)\,, \;\;\; M^2_{B_H} = \frac15 {g'}^2 f^2 \left(1+\frac{1}{5}{r}^2\right)\,, \;\;\text{with} \;\; r\equiv \frac{f'}{f}\,. \label{gaugeMasses}
\end{align}
We recognize the orthogonal linear combinations,  
\begin{align}
W^a = \frac{1}{\sqrt{2}}(W_1^a+W_2^a)\,,\;\; B = \frac{1}{\sqrt{2}}(B_1+B_2) \,,
\end{align}
as the SM gauge fields. 
\subsection{Goldstone sector} 
In addition to the complex Higgs doublet $H$ and the charged triplet $\Phi$, the Goldstone sector includes additional physical states: a real singlet $s$ and a real triplet $\varphi \equiv \frac12 \varphi_a \sigma^a$, defined as the following linear combinations
\begin{align}
s = c_0 \pi_0+s_0 \phi_0 \,,\;\;\; \varphi_a = c_{3} \tau_a-s_{3} \pi_a\,, \label{physPNGBs}
\end{align}
with the mixing angles
\begin{align}
s_0 = \sqrt{1-c_0^2} \equiv \frac{r}{\sqrt{5+r^2}}\,, \;\;  c_3 = \sqrt{1-s_3^2} \equiv \frac{r}{\sqrt{1+r^2}}\,.  \label{NGBangles}
\end{align}
The orthogonal linear combinations,
\begin{align}
G_0 = -s_0 \pi_0+c_0 \phi_0 \,,\;\;\; G^a = s_{3} \tau_a+c_{3} \pi_a\,, \label{eatenPNGBs}
\end{align}
are "eaten" by the heavy gauge fields and removed from the spectrum in the unitary gauge. 
\subsection{Matter sector} 
The top Yukawa generates the largest quadratically divergent contribution to the Higgs mass, therefore we limit our discussion to the third quark family.  The terms in the top sector must respect enough of the global symmetries in order for the Higgs mass to be protected from 1-loop quadratically divergent contributions. This mechanism is usually referred to as "collective" symmetry breaking. 
In order to respect these symmetries we enlarge the multiplets introduced in \Eq{psi1psi2} and introduce top partners. The quadratically divergent contribution to the Higgs mass from these top partners would eventually cancel out with the top contribution. We start by introducing left-handed Weyl fermions. We embed the doublets $\psi_{1,2}$ with the singlets $\chi_{1,2}$ (the top partners) in incomplete $SU(5)$ multiplets 
\begin{align}
\Psi_1 = \begin{pmatrix}
\psi_1 \\ \chi_1 \\0
\end{pmatrix}_{\overline{\bf{5}}}\,,
\;\;
\Psi_2 = \begin{pmatrix}
0 \\ \chi_2 \\ \psi_2
\end{pmatrix}_{\bf{5}}\,.
\end{align}
Under $T$-parity, 
\begin{align}
\Psi_1 \to \Omega \Sigma_0 \Psi_2\,,
\end{align}
or equivalently
\begin{align}
\psi_1 \leftrightarrow \psi_2 \,, \;\; \chi_1 \leftrightarrow -\chi_2\,.
\end{align}
We introduce 3 right-handed singlets denoted by  $\tilde{t}_R, \tau_{1,2}$. Under $T$-parity,
\begin{align}
\tilde{t}_R \leftrightarrow \tilde{t}_R\,, \;\; \tau_1 \leftrightarrow \tau_2\,.
\end{align}
The top Yukawa is given by~\cite{ArkaniHamed:2002qy,Low:2004xc}
\begin{align}
\Ltop &= \frac{\lambda_1 f}{2}\left(  {\overline \Psi_1}_i O_i+   (\overline \Psi_2  \Omega \Sigma_0)_i \tilde O_i \right) \tilde t_R+\frac{\lambda_2 f}{\sqrt{2}}\left( \overline{\chi}_1 \tau_1 - \overline{\chi}_2 \tau_2 \right)+\text{h.c}\,, \nonumber
\\O_i &\equiv  \epsilon_{ijk}\Sigma_{j4}\Sigma_{k5}\,,\;\; \tilde{O}_i \equiv  \epsilon_{ijk}\tilde\Sigma_{j4}\tilde\Sigma_{k5}\,.
\end{align}
 $\tilde\Sigma$ is defined in \Eq{sigmatilde}. The indices $i,j,k$ are summed over $1,2,3$. We define the $T$-parity eigenstates 
 \begin{align}
 \Psi_+ &= \frac{1}{\sqrt{2}}\left(\Psi_1  +\Omega \Sigma_0 \Psi_2 \right) \equiv \begin{pmatrix}
\sigma_2 Q_L \\
\chi_+ \\
0
 \end{pmatrix}\,, \;\;  \Psi_- = \frac{1}{\sqrt{2}}\left(\Psi_1  -\Omega \Sigma_0 \Psi_2 \right) \equiv \begin{pmatrix}
\sigma_2 \psi^-_L  \\
T^-_L  \\
0
 \end{pmatrix}\,, 
\end{align}
 with
\begin{align}
 Q_L = \begin{pmatrix} \tilde{t}_L \\ b_L \end{pmatrix} =  \frac{1}{\sqrt{2}}\sigma_2(\psi_1+\psi_2)\,, \;\; \psi^-_L =  \frac{1}{\sqrt{2}}\sigma_2(\psi_1-\psi_2)\,.
\end{align} 
The singlet $T$-parity eigenstates are defined as
 \begin{align}
\chi_+ &= \frac{1}{\sqrt{2}}\left( \chi_1 - \chi_2 \right)\,, \;\; \tau_+ = \frac{1}{\sqrt{2}}\left( \tau_1 + \tau_2 \right)\,, \;\;  T^-_L = \frac{1}{\sqrt{2}}\left( \chi_1 + \chi_2 \right)\,, \;\; T^-_R = \frac{1}{\sqrt{2}}\left( \tau_1 - \tau_2 \right)\,.
 \end{align}
Note that the $T$-even fields, and in particular $\tilde{t}_L,\tilde{t}_R$, are not the mass eigenstates (hence the tilde).
After the Higgs field acquires its VEV, $\left< H \right> = \frac{1}{\sqrt{2}}( 0,v)^T$, we find the following mass matrix for the $T$-even fermions
\begin{align}
\Ltop   \ni f\begin{pmatrix}
 \overline{\tilde{t}}_L & \overline{\chi}_+
\end{pmatrix}
\begin{pmatrix}
  \frac{\lambda_1 s_v }{2}&0\\
\frac{\lambda_1 (1+c_v) }{2\sqrt{2}}& \frac{\lambda_2 }{\sqrt{2}}
\end{pmatrix} \begin{pmatrix}
\tilde{t}_R 
\\
\tau_+
\end{pmatrix}+\hc\,.
\end{align}
We denoted
\begin{align}
s_v = \sin \sqrt{2\xi }\,, \;\;\; c_v = \cos \sqrt{2\xi }\,, \;\;\; \xi \equiv \frac{v^2}{f^2}\,.
\end{align}
The physical basis is given by
\begin{align}
\begin{pmatrix}
t_L \\
T^+_L
\end{pmatrix}
=
\begin{pmatrix}
c_L & -s_L \\
s_L &c_L
\end{pmatrix}
\begin{pmatrix}
\tilde{t}_L \\
\chi_+
\end{pmatrix}\,,\;\; 
\begin{pmatrix}
t_R \\
T^+_R
\end{pmatrix}
=
\begin{pmatrix}
c_R & -s_R \\
s_R & c_R
\end{pmatrix}
\begin{pmatrix}
\tilde{t}_R \\
\tau_+
\end{pmatrix}\,,
\end{align}
with $\sin \theta_{L/R} \equiv s_{L/R}$ and  $\cos \theta_{L/R} \equiv c_{L/R}$. The mixing angles are given by~\cite{Hubisz:2005tx}
\begin{align}
\theta_L &= \frac12 \tan^{-1} \left( \frac{2\sqrt{2} \lambda_1^2 s_v (1+c_v)}{4\lambda_2^2+(1+c_v)^2\lambda_1^2-2\lambda_1^2 s_v}\right)\,,
\\
\theta_R &= \frac12 \tan^{-1} \left( \frac{4 \lambda_1 \lambda_2  (1+c_v)}{4\lambda_2^2-\lambda_1^2(2s_v^2+(1+c_v)^2)}\right)\,.
\end{align}
 The masses at leading order in $\xi$ are
\begin{align}
m_t^2 = \frac{1}{\sqrt{2}} \left(\frac{\lambda_1\lambda_2}{\sqrt{\lambda_1^2+\lambda_2^2}}\right) \sqrt{\xi} f \,, \;\;\; m_{T^+} =  \frac{\sqrt{\lambda_1^2+\lambda_2^2}}{\sqrt{2}}f\,.
\end{align}
The top Yukawa coupling at leading order in $\xi$ is therefore
\begin{align}
y_t = \frac{\lambda_1\lambda_2}{\sqrt{\lambda_1^2+\lambda_2^2}}\,.
\end{align}
We shall keep $\lambda_2$ as a free parameter and fix $\lambda_1$ to produce the correct top Yukawa $y_t \approx 1$. The mixing angles at leading order in $\xi$ are
\begin{align}
s_L =  \frac{\lambda_1^2}{\lambda_1^2+\lambda_2^2}\sqrt{\xi}  = \left(\frac{y_t}{\lambda_2}\right)^2\sqrt{\xi} \,, \;\;\; s_R = \frac{\lambda_1}{\sqrt{\lambda_1^2+\lambda_2^2}} = \frac{y_t}{\lambda_2}\,.
\end{align}
For the $T$-odd sector we must introduce a mass term for the doublet similar to the term in \Eq{quasi_kappa}. We introduce a RH doublet $\psi_R^-$ transforming non-linearly under $[SU(2)\times U(1)]_L \times [SU(2)\times U(1)]_R$ according to the CCWZ formalism. $\psi_R^-$ is odd under $T$-parity
\begin{align}
\psi_R^- \to -\psi_R^-\,.
\end{align}
The mass term is given by~\cite{Pappadopulo:2010jx}
\begin{align}
\Lagr_{\kappa} =  \frac{\kappa f}{\sqrt{2}}\left( \overline \psi_1 \sigma_2 e^{\frac{i}{f'}\Pi_X} - \overline \psi_2 \sigma_2 e^{-\frac{i}{f'}\Pi_X}  \right ) \psi_R^- +\hc\,. \label{Lkappa}
\end{align}
Our spectrum contains a $T$-odd singlet $T^-$ and a $T$-odd doublet $\psi^-$ with the following masses
\begin{align}
m_{T^-} = \frac{\lambda_2}{\sqrt{2}} f\,, \;\;\; m_{\psi^-} =\kappa f\,.
\end{align}
Lastly, the explicit form of the kinetic terms can be found in \App{modelApp}. 
\section{Scalar potential} \label{ScalarPotentialSec}
At tree level, the pNBG's interact only through derivative interactions and their classical potential vanishes. The gauge and top sector couplings explicitly break the global symmetry.  The classical scalar potential is radiatively generated from fermion and gauge loops. At 1-loop the fermion and gauge loops contributions are given by~\cite{Coleman:1973jx}
\begin{align}
V_{f}(H,\Phi,s,\varphi) =&  -  \frac{N_c}{8 \pi^2} \Lambda^2a_1\,\text{Tr}\left[ M_{f}M^\dagger_{f} \right]  -  \frac{N_c}{16 \pi^2}a_2\text{Tr}\left[ M_{f}M^\dagger_{f} M_{f}M^\dagger_{f}  \log \left(\frac{M_{f}M^\dagger_{f} }{\Lambda^2 } \right)\right]\,, \label{fermionCW}
\\
V_{\text{\tiny V}} (H,\Phi,s,\varphi) =&\frac{3}{32 \pi^2}  \Lambda^2 a_3\,\text{Tr}\left[ M^2_{\text{\tiny V}} \right]+ \frac{3}{64 \pi^2}a_4\text{Tr}\left[   M^2_{\text{\tiny V}}M^2_{\text{\tiny V}}  \log \left(  \frac{M^2_{\text{\tiny V}} }{\Lambda^2 }\right)\right]\,, \label{gaugeCW}
\end{align}
respectively. $M_{f}(H,\Phi,s,\varphi)$ and $M^2_{\text{\tiny V}}(H,\Phi,s,\varphi)$ are the fermion and gauge bosons mass matrices in the background of the pNGB's. The $a_i$ parameters with $i=1,..,4$ are unknown $O(1)$ numbers originating from unknown UV contributions to these operators. $ \Lambda \sim 4 \pi f $ is the cutoff scale of the theory. Expanding the scalar potential $V = V_{f} +V_{\text{\tiny V}} $ in the NGB fields, we find that
\begin{align}
V   = m_{\Phi}^2 \text{Tr}[\Phi^\dagger\Phi] -\mu^2 |H|^2 + {m}_\varphi^2 \text{Tr}[\varphi^2]+\lambda |H|^4 + \ldm s^2 |H|^2 + \lambda_{\varphi}s H^\dagger \varphi  H+...\,.
\label{scalarPotentialEq}\end{align}
We have omitted additional radiatively generated operators that are inconsequential for the upcoming discussions. A detailed analysis of the symmetries of the scalar potential of this model can be found in \App{scalarPotential}.  In this section we summarize the most important features of the scalar potential.
\\ \\
The mass of the charged triplet $\Phi$ is quadratically divergent,
\begin{align}
m_{\Phi}  \sim \text{a few TeV}\,.
\end{align}
We consider energy scales well below $m_{\Phi}$.  We remove $\Phi$ from our spectrum by integrating it out. Due to $T$-parity, integrating out $\Phi$ at tree-level does not influence any of the couplings explicitly written in the scalar potential of \Eq{scalarPotentialEq}.
Like $m_{\Phi}^2$, the Higgs quartic $\lambda$ is generated by 1-loop quadratically divergent diagrams.
 \\ \\
The rest of the operators in \Eq{scalarPotentialEq}, including the Higgs mass term $\mu^2$, are generated through logarithmically divergent loops, and as such they exhibit a mild dependence on the UV cutoff scale. The explicit calculations, found in \App{scalarPotential}, give us an order of magnitude estimation for the IR contribution to these operators at 1-loop. However quadratically divergent 2-loop diagrams as well as UV contributions can have comparable effects on these operators. Therefore we do not presume to be able to predict  these couplings accurately in terms of the fundamental parameters of this model. In this work we treat the couplings in \Eq{scalarPotentialEq} as free parameters, except $\mu^2$ and $\lambda$ which are already fixed by experiment. Our goal is to allow the free parameters to take values that are reasonable in light of the approximation given by the 1-loop IR contribution, and state explicitly when this is not the case. 
\\ \\
In addition to $ {m}_\varphi^2,\lambda_{\text{\tiny DM}} ,\lambda_{\varphi}$, we must introduce a mass term for the singlet $s$. The singlet remains massless at 1-loop, and  a mass for $s$ is generated at the 2-loop level. We take the pre-EWSB mass term of the singlet, denoted as $\tilde m_s^2$, as a free parameter as well. The sizes and ranges of $ {m}_\varphi^2,\lambda_{\varphi},\tilde m_s^2,\lambda_{\text{\tiny DM}}$ are dictated by the DM phenomenology and are discussed in \Sec{dmPheno}.
\section{LHC phenomenology}\label{lhcPhen}
\subsection{$T$-even singlet $T^+$}
The $T$-even singlet is responsible for cancelling the quadratically divergent top loop contribution to the Higgs mass, hence it is the standard top partner predicted by composite Higgs models. It can be doubly produced at the LHC via QCD processes, as well as singly produced with an associated third generation quark through the following EW interactions
\begin{align}
\Lagr \ni \frac{g}{2}C_{bW} \bar{T}_L^+ \slashed{W} b_L + \frac g2C_{tZ} \bar{T}_L^+ \slashed{Z} t_L+\hc\,.
\end{align}
In this model,
\begin{align}
C_{bW} &= \sqrt{2} s_L \approx \frac{\sqrt{2 \xi}}{\lambda_2^2} \approx 0.35\left(\frac{1}{\lambda_2}\right)^2 \left( \frac{1~\TeV}{f}\right) \,, \nonumber
\\
C_{tZ} &= \frac{ s_L c_L }{ c_W  } \approx  \frac{\sqrt{ \xi}}{ c_W \lambda_2^2} \approx 0.28\left(\frac{1}{\lambda_2}\right)^2 \left( \frac{1~\TeV}{f}\right)\,. \label{prodCoef}
\end{align}
\subsubsection*{Decay modes}
\begin{figure}
\center
\includegraphics[width=0.45\textwidth]{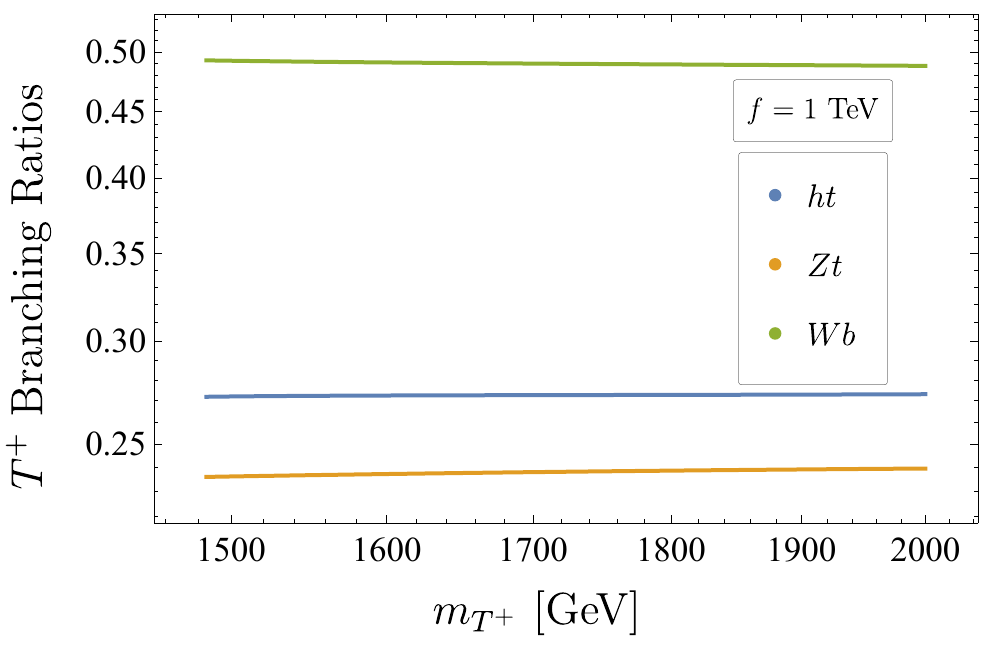}
\includegraphics[width=0.45\textwidth]{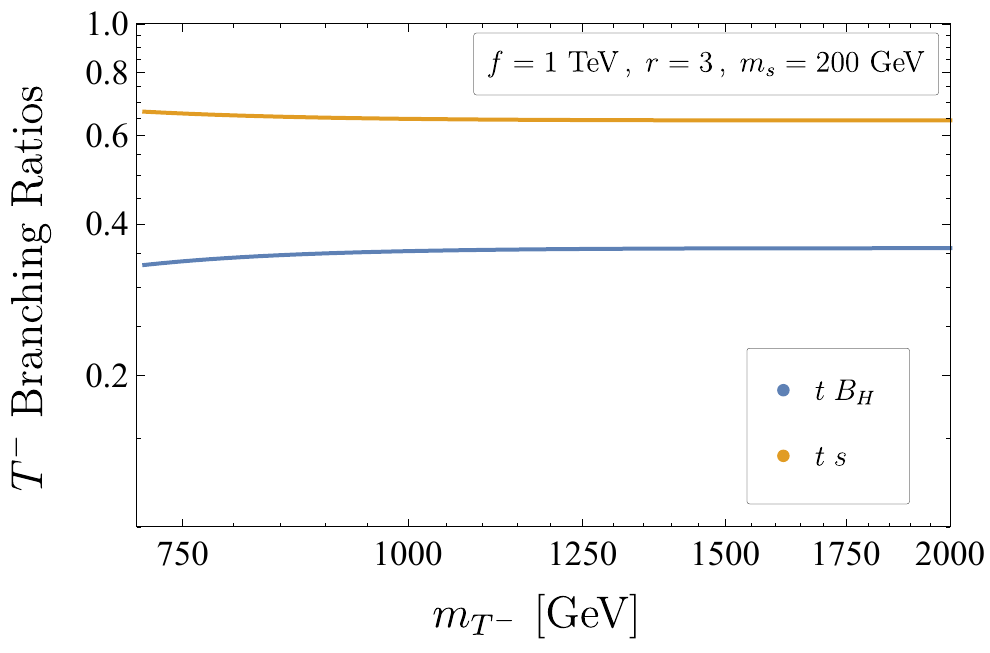}
\caption{Left panel : Numeric results for branching ratios of $T^+$ for $f=1$~TeV. Right Panel: Numeric results for branching ratios of $T^-$ for $f=1$~TeV, $r=3$ and $m_s=200$~GeV. The mass of $B_H$ is a function of $f,r$, in this case $m_{B_H} = 270~$GeV.  }
\label{brs}
\end{figure}

We consider the limit $m_{T^+} \gg m_H,m_W,m_Z$. In this regime EWSB effects are negligible and we can formally take $\xi \to 0$. The dominant decays of  $T^+$ are to the physical Higgs or to the longitudinal components of the SM gauge bosons with an associated third generation quark, in accordance with the equivalence theorem. We can parameterize the Higgs field in a general $R_\xi$ gauge using these would-be longitudinal components as
\begin{align}
H = \begin{pmatrix}
\phi^+ \\
\frac{1}{\sqrt{2}}(v+ h+i \phi_0)
\end{pmatrix}\,.
\end{align}
The relevant interactions between the Higgs doublet and $T^+$ are
\begin{align}
\Ltop  \ni &  -\frac{1}{\sqrt{2}}\lambda_1\sin\theta_R \left( \overline{t}_L(v+ h+i \phi_0)-\sqrt{2}\overline{b}_L\phi^+\right)T^+_R+\text{h.c}
\end{align}
Predicting that in the high energy limit,
\begin{align}
\text{Br}[T_+ \to h\; t] : \text{Br}[T_+ \to Z\; t] : \text{Br}[T_+ \to W^+\;b] = 1 : 1 : 2\,.
\end{align}
The exact branching ratios $T^{+}$ including EWSB and phase space effects can be found on the left panel in \Fig{brs}. 
\subsubsection*{LHC searches}
\emph{Single production}: $T_+$ can be singly produced at the LHC in association with a third generation quark.  A recent search from CMS~\cite{CMS:2017oef} looked for $(T^+ \to Z\;t) bq$ with a fully leptonic $Z$ decay. The search places a lower bound on the mass of the singlet LH Top partner at $1.2$~TeV, assuming negligible width and BR$[T^+\to Z t]=0.25$. The bound strongly relies on a model-dependent production cross-section, which in term depends on the coefficients of \Eq{prodCoef}. In the CMS search the coupling is fixed at $C_{bW}=0.5$. 
Conservatively we consider the $m_{T^+}>1.2$~TeV bound at face value, although we expect a smaller value for $C_{bW}$, as can be seen in \Eq{prodCoef}. $C_{bW}$ is further suppressed for $\lambda_2>1$, which is the region in parameters space that, as we later show, is consistent with the LHC constraints on the $T$-odd top partners masses.  The mass of the $T$-odd singlet is bound from below to be $m_{T^+}  > \sqrt{2} f$. The lower bound of $1.2$~TeV can be trivially satisfied by taking $f>850$~GeV. 
\\\emph{Double production}: $T_+$ can also be doubly-produced via QCD processes. A recent search from ATLAS~\cite{ATLAS:2016btu} looked for a pair produced top partners in a range of final states,  assuming that at least one of the top partner decays to $ t h$. The quoted nominal bound of the singlet top partner is 
\begin{align}
m_{T^+} > 1.02~\text{TeV}\,. \label{TevenBound}
\end{align}
This bound can be satisfied by taking $f> 700$~GeV. 
\subsection{$T$-odd singlet $T^-$}
The phenomenology of the $T$-odd singlet resembles that of a stop squark with conserved R-parity. It can be doubly produced at the LHC via QCD processes, and consequently decay to tops and missing energy. \subsubsection*{Decay modes}
We consider the limit $m_{T^-} \gg m_s,m_{B_H},m_t$.  $T^{-}$ couples to the singlet $\phi_0$ of the original $\frac{SU(5)}{SO(5)}$ coset. In a general $R_\xi$ gauge, $\phi_0$  is composed of the physical singlet and the would-be longitudinal component of $B_H$,
\begin{align}
\phi_0 = s_0 s + c_0 G_0\,.
\end{align}
The relevant interactions are
\begin{align}
\Ltop \ni   i\lambda_1\sqrt{\frac{2}{5}}   \left( \phi_0\bar{T}^{-} t_R \right)+\hc= i\lambda_1\sqrt{\frac{2}{5}}   \left( s_0\; s\;\bar{T}^{-} t_R +c_0\;G_0\;\bar{T}^{-} t_R\right)+\hc\,.
\end{align}
Leading to the simple prediction in the high energy limit
\begin{align}
\Gamma(T^- \to s\; t) : \Gamma(T^- \to B_H t) = \left( \frac{s_0}{c_0}\right)^2 = \frac{r^2}{5}
\end{align}
The exact branching ratios of $T^{-}$ including EWSB and phase space effects can be found on the right panel in \Fig{brs}.
\subsubsection*{LHC searches}
We performed a simple recast of recent stop bounds by accounting for the enhanced production cross section of the fermionic $T^-$ relative to the scalar stop squark case. We would like to account for the presence of the $T$-odd doublet, which contributes to the same final states as $T^-$.  We postpone the derivation of these bounds to \Sec{psiminussection}.
\subsection{$T$-odd doublet $\psi^-$} \label{psiminussection}
The phenomenology of the $T$-odd doublet resembles that of a mass-degenerate stop and sbottom squarks with conserved R-parity. The upper (lower) component up $\psi^-$ can be doubly produced at the LHC via QCD processes, and consequently decay to tops (bottoms) and missing energy.
\subsubsection*{Decay modes}
\begin{figure}
\center
\includegraphics[width=0.49\textwidth]{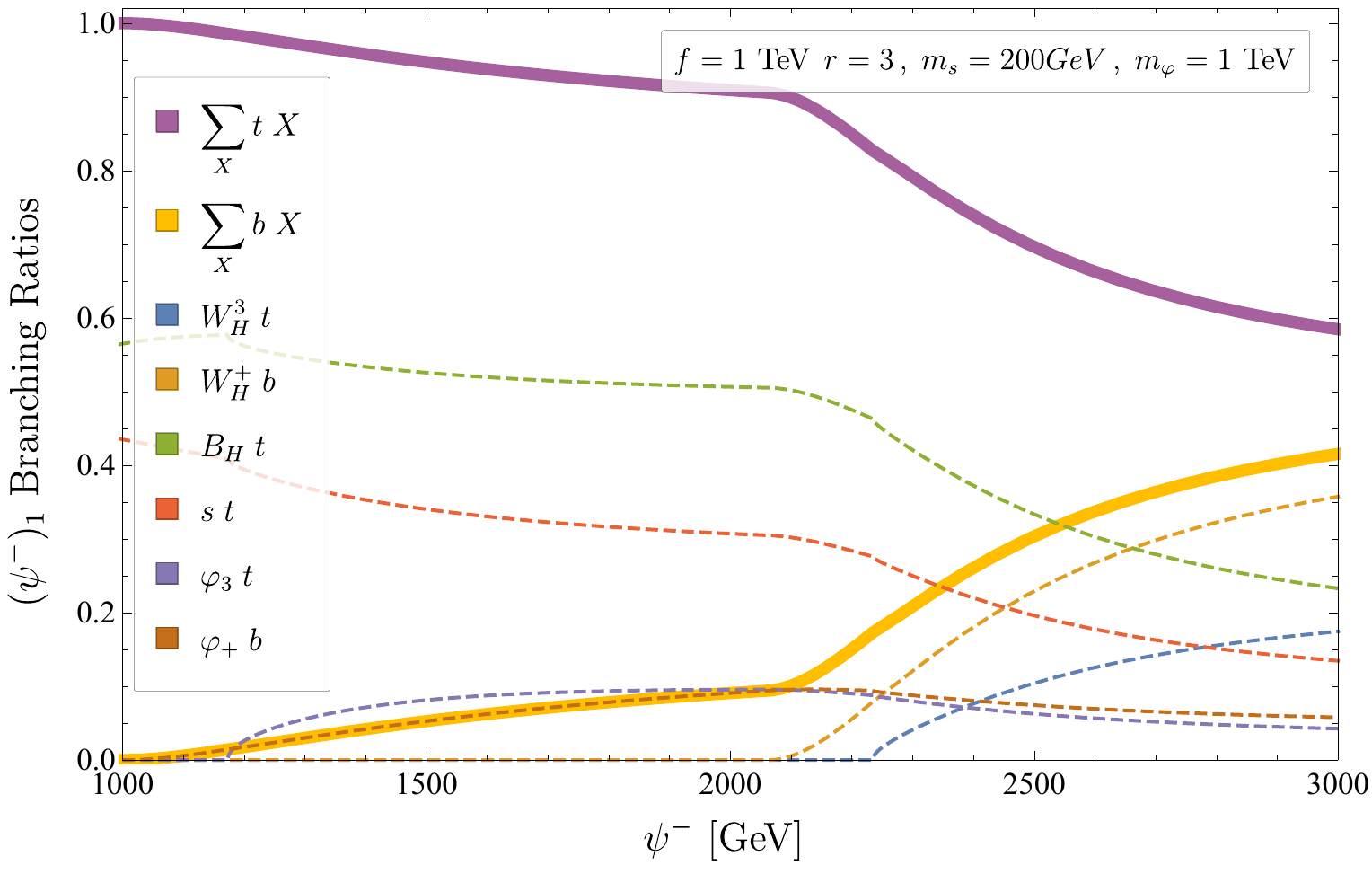}
\includegraphics[width=0.49\textwidth]{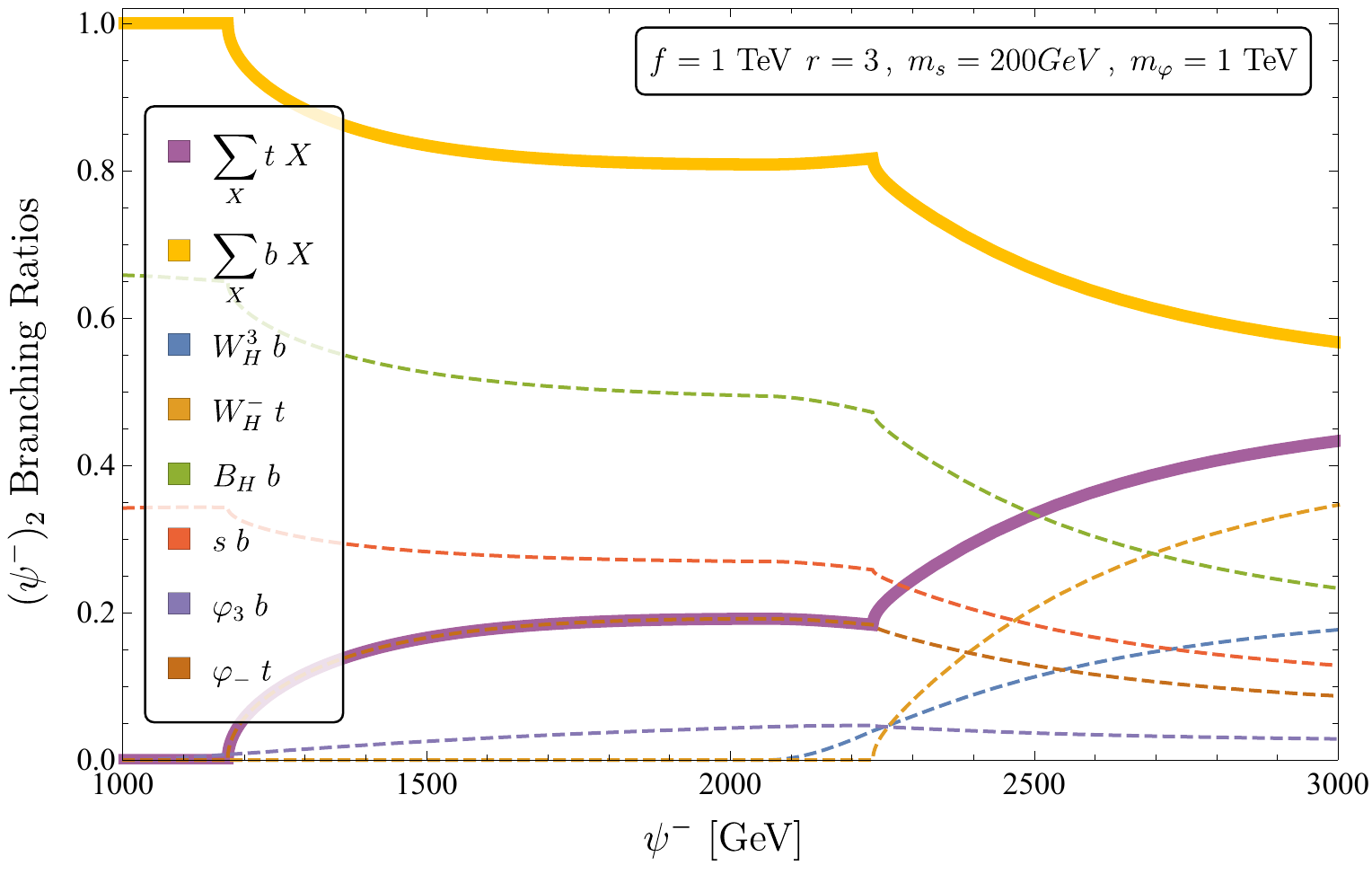}
\caption{Numeric results for the branching ratios of the upper (lower) component of $\psi_-$ presented in the left (right) panel,  with $f~=~1~\text{TeV},r~=~3, m_s=200~\text{GeV},m_{\varphi}=1~\text{TeV}$ and $\lambda_2=2.5$. The masses of the heavy gauge boson are fixed at $m_{B_H} = 270$~GeV and $m_{W_H} = 2.1$~TeV. The dashed colored lines indicate the branching ratios to the different exclusive final states. The solid thick lines indicate the sum of branching ratios  with either a top (purple curve) or a bottom (yellow curve) at the final state.  }
\label{psiminusbrs}
\end{figure}

We consider the limit where $m_{\psi^-} \gg m_{B_H},m_{W_H},m_s,m_\varphi$. In a general $R_\xi$ gauge, we can express our original pNGB's in terms of the physical pNGB's and the would-be longitudinal modes of the heavy gauge fields defined in \Eqs{physPNGBs}{eatenPNGBs},
 \begin{align}
\begin{pmatrix}
\tau^a \\
\pi^a
\end{pmatrix}
=
\begin{pmatrix}
c_3 & s_3 \\
-s_3 & c_3 
\end{pmatrix}
\begin{pmatrix}
\varphi^a \\
G^a
\end{pmatrix} 
\,, \;\; 
\begin{pmatrix}
\pi^0 \\
\phi_0 
\end{pmatrix}
=
\begin{pmatrix}
c_0 & -s_0 \\
s_0 & c_0 
\end{pmatrix}
\begin{pmatrix}
s \\
G^0
\end{pmatrix} 
\,. \end{align}
with the mixing angles $c_0,c_3,s_0,s_3$ defined in \Eq{NGBangles}.
The relevant interaction in the $\xi \to 0$ limit originate from $\Lagr_{\kappa}$. For $(\psi^-_R)^1$,
\begin{align}
\Lagr_{\kappa} \ni \frac{i\kappa }{2}  \frac1r\left[ (c_0 s-s_0 G_0)  \overline{t}_L +(-s_3 \varphi_3+c_3 G_3 ) \overline t_L
+\sqrt{2}(-s_3 \varphi^-+c_3 G^- )\overline{b}_L\right](\psi^-_R)^1\,,
\end{align}
and similarly for $(\psi_-^R)^2$,
\begin{align}
\Lagr_{\kappa} \ni \frac{i\kappa }{2} \frac1r \left[ (c_0 s-s_0 G_0)  \overline{b}_L -(-s_3 \varphi_3+c_3 G_3 ) \overline b_L
+\sqrt{2}(-s_3 \varphi^++c_3 G^+ )\overline{t}_L\right](\psi^-_R)^2\,.
\end{align}
In the high energy limit
\begin{align}
&\text{Br}[\psi^- \to q \;s]  =  \frac{c_0^2}{4}\,,\;\;\; \text{Br}[\psi^- \to q \;G_0] = \frac{s_0^2}{4}\,,
\\&\text{Br}[\psi^- \to q \;\varphi_3]  = \frac12 [\psi^- \to q \;\varphi_\pm]  = \frac{s^2_3}{4}\,,
\\&\text{Br}[\psi^- \to q \;G_3]  = \frac12 [\psi^- \to q \;G_\pm ] =\frac{c^2_3}{4}\,.
\end{align}
with the final state with $q= \{b,t\}$ depending on the electric charge of the initial state. The exact branching ratios for $\psi^{-}$ including EWSB and phase space effects can be found in \Fig{psiminusbrs}.
\subsubsection*{LHC searches}
\begin{figure}
\center
\includegraphics[width=0.4\textwidth]{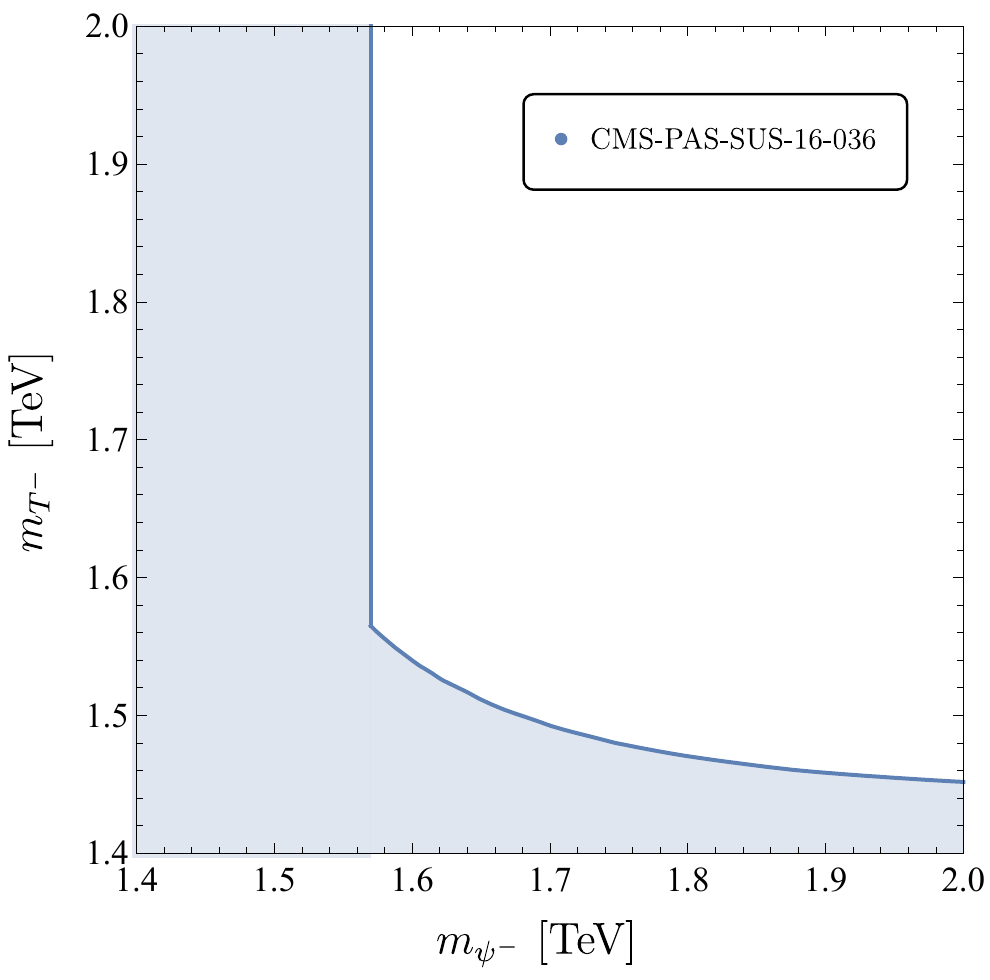}
\includegraphics[width=0.4\textwidth]{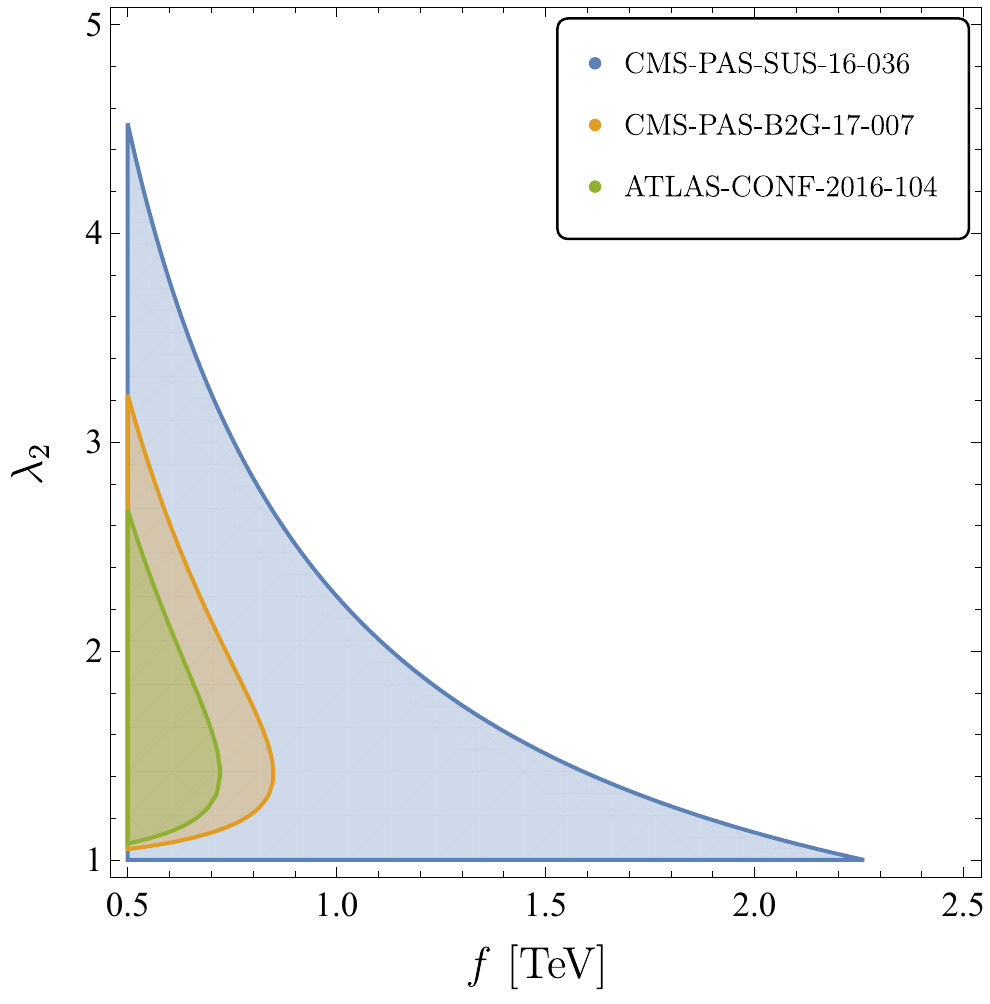}
\caption{Left Panel: Exclusion limits (blue region) in the $(m_{\psi^-},m_{T^-})$ plane, using recasted limits from the CMS SUSY search of~\Ref{CMS:2017kmd}. We impose the condition of \Eqs{cond1}{cond2}, assuming branching ratios of 100\%. Right Panel: Exclusion limits in the $(f,\lambda_2)$ plane using \Ref{CMS:2017kmd} (blue region, using the bound from \Eq{lambda2lowerbound}), \Ref{CMS:2017oef} (orange region) and \Ref{ATLAS:2016btu} (green region).}
\label{toddrecastplot}
\end{figure}

The $T$-odd sector contains two top-like and one bottom-like fermions.  We perform a recast of recent bounds on stop and sbottom masses by accounting for the enhanced production cross section of a fermionic colored top partner. The quoted bounds in \Ref{CMS:2017kmd} for the stop and sbottom masses are
\begin{align}
m_{\tilde t} \geq 1070~\text{GeV}\,, \;\;\; m_{\tilde b} \geq 1175~\text{GeV}\,,
\end{align}
respectively. We denote the QCD pair production cross section at $\sqrt{s}= 13$~TeV for a spin $s$ coloured particle with mass $M$ as $\sigma^s_{\text{\tiny pair}}(M)$.  We require that
\begin{align}
&\text{I} : \sigma^0_{\text{\tiny pair}}(1070~\text{GeV}) \geq \sigma^{1/2}_{\text{\tiny pair}}(m_{\psi^-})\times \text{BR}[(\psi_-)_1 \to t+\text{MET}]+ \sigma^{1/2}_{\text{\tiny pair}}\left(m_{T^-}\right) \label{cond1}\;\;\; \text{and}
\\
&\text{II} : \sigma^0_{\text{\tiny pair}}(1175~\text{GeV}) \geq \sigma^{1/2}_{\text{\tiny pair}}(m_{\psi^-})\times\text{BR}[(\psi_-)_2 \to b+\text{MET}]\,,\label{cond2}
\end{align}
with $m_{\psi^-} = \kappa f$ and $m_{T^-} = \frac{\lambda_2 f}{\sqrt{2}}$ the masses of the $T$-odd doublet and $T$-odd singlet top partners respectively. We use $\sigma^0_{\text{\tiny pair}}(M)$ reported by the CMS collaboration~\cite{Borschensky:2014cia} and $\sigma^{1/2}_{\text{\tiny pair}}(M)$ calculated using HATHOR~\cite{Aliev:2010zk}. The combination of I+II in the $(m_{\psi^-},m_{T^-})$ plane is plotted in the left panel of \Fig{toddrecastplot}. We conservatively assume all the branching ratios to be $100\%$. We thus obtain the following lower bounds on the $T$-odd fermion masses
\begin{align}
m_{\psi^-},m_{T_-} > 1.6~\text{TeV}\,.\label{massBound}
\end{align}
The combination of all LHC constrains in the $(f,\lambda_2)$ plane is shown in the right panel of \Fig{toddrecastplot}.  We summarize the constraints for the couplings for a for a given $f$, 
\begin{align}
\frac{1.6~\text{TeV}}{f}<\kappa < 4 \pi\,, \quad  \text{Max}\left[1,\frac{2.3~\text{TeV}}{f}\right]<\lambda_2 < 4 \pi\,. \label{lambda2lowerbound}
\end{align}
\section{Electroweak precision tests} \label{ewpt}
The main contributions to electroweak precision observables are unaffected by the extended coset structure and briefly review updated results available in the literature. 
The mixing in the left-handed sector generates a correction to the T oblique parameter due to loops of the $T$-even singlet $T^+$~\cite{Hubisz:2005tx}
\begin{align}
T_{T^+} &= T_\text{\tiny SM}  \,{s_L}^2\,\,\left[\frac{{s_L}^2}{x_t}-2+{s_L}^2-\frac{2{s_L}^2}{1-x_t}\log x_t \right]\,,  \label{TparTeven}
\end{align}
with
\begin{align}
T_\text{\tiny SM}  = \frac{3}{16\pi}\,\frac{1}{s_w^2c_w^2}\,\frac{m_t^2}{m_Z^2}\approx 1.24\,, \;\; x_t &\equiv \frac{m_t^2}{m_{T^+}^2} \approxeq \left(\frac{\lambda_2^2-1}{\lambda_2^4}\right)\xi 
\,,
\end{align}
and
\begin{align}
s_L \equiv \sin \theta_L \approxeq \sqrt{\frac{x_t}{\lambda_2^2-1}} \,. \label{sLxT}
\end{align}
We express $T_{T^+}$ in terms of $x_t$ using \Eq{sLxT}. In light of the LHC constrains on the $T$-even top partner mass of \Eq{TevenBound}, we expect $x_t \leq 0.03 \ll 1$. We therefore expand \Eq{TparTeven} to leading order in $x_t$:
\begin{align}
T_{T^+} &\approx T_\text{\tiny SM}\left( \frac{ x_t } {\lambda_2^2-1} \right)\left(2\log\frac{1}{x_t}+ \left[\frac{1}{\lambda_2^2-1}\right]-2\right)\\& = T_\text{\tiny SM} \left( \frac{  \xi }{\lambda_2^4} \right)\left(2\log\left[ \frac{\lambda_2^4}{(\lambda^2_2-1)\xi}\right]+ \left[\frac{1}{\lambda_2^2-1}\right]-2\right) \,.
\end{align}
An additional contribution to the T parameter is due to loops of $T$-odd heavy gauge bosons. The correction is proportional to the mass splitting after EWSB,
\begin{align}
\Delta m_{W_H}^2 \equiv  m_{W^3_H}^2-m_{W^\pm_H}^2 = \frac12 f^2 g^2 \sin^4 \left(\sqrt{\frac{\xi}{2}} \right)\,,
\end{align}
neglecting corrections of order $O(g'^2)$. The $T$-odd gauge loops generate the following correction to the T parameter~\cite{Hubisz:2005tx}
\begin{align}
T_{W_H} = - \frac{9}{16 \pi c_w^2 s_w^2 M_Z^2} \Delta m_{W_H}^2 \log
  \left(\frac{\Lambda^2}{f^2 g^2 (1+r^2)}\right) = - \frac{9}{16 \pi s_w^2 }  \xi \log
  \left(\frac{\Lambda}{f g \sqrt{1+r^2}}\right). \label{TWH}
\end{align}
This correction is $\lambda_2$ independent, and becomes the dominant one for higher values of $\lambda_2$ as $T_{T^+}\to0$. We assume that the UV contributions to these loop processes are sub-leading with respect to the log-enhanced IR contribution.
\\ \\
Let us mention that the oblique $S$ and $U$ parameters also receive corrections due to the mixing the LH fermion sector. As noted in \Ref{Hubisz:2005tx},  the size of these corrections are an order of magnitude smaller than the correction to the $T$ parameter and are therefore sub-leading. 
Additionally, the $Z \bar{b}_L b_L$ vertex receives corrections due to $T_+$ loops~\cite{Hubisz:2005tx}
\begin{align}
\delta g_L^{Z\bar{b}b}=\frac{g}{c_w}\frac{\alpha}{8\pi s_w^2} \frac{m_t^4}{m_W^2 m_{T^+}^2}
\left(\frac{1}{\lambda_2^2-1}\right) \log \frac{m_{T^+}^2}{m_t^2}\,,
\end{align}
with $\delta g_L^{Z\bar{b}b} \equiv  g_L^{Z\bar{b}b} -  g_{L\;\SM}^{Z\bar{b}b}$ and $g_{L\;\SM}^{Z\bar{b}b} = -\frac12 + \frac{s^2_w}{3}$.  We constrain the parameters of the model using the results of~\Ref{deBlas:2016ojx}, namely
\begin{align}
T &=  0.12 \pm 0.07\,,
\\
\delta g_L^{Z\bar{b}b}&= 0.002 \pm 0.001\,.
\end{align}
\begin{figure}
\center
\includegraphics[width=0.6\textwidth]{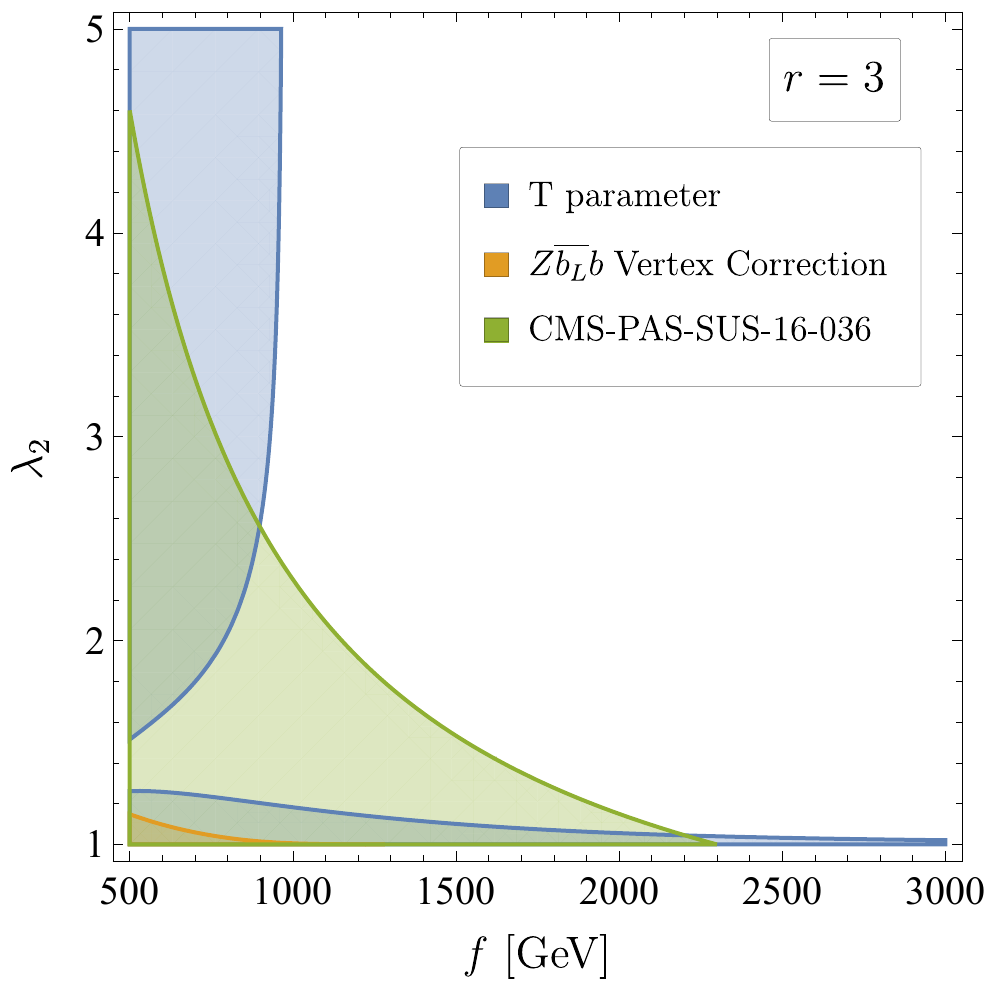}
\caption{Combined EWPT and LHC exclusion regions in the $(f,\lambda_2)$ plane, for $r=3$ and $\Lambda = 4\pi f$. The EWPT exclusion regions due to T-parameter (blue region) and $\delta g_L^{Z\bar{b}b}$ (orange region) are plotted at the $3\sigma$ level using the results of~\Ref{deBlas:2016ojx}, $T~=~0.12~\pm~0.07$ and $\delta g_L^{b\bar{b}}~=~0.002~\pm~0.001$. The LHC exclusion (green region) is due to \Ref{CMS:2017kmd} using the lower bound of \Eq{lambda2lowerbound}. }
\label{EWPT}
\end{figure}
The combinations of the EWPT and LHC constraints are plotted in ~\Fig{EWPT}.  For $f\lesssim1.5$~TeV values of $\lambda_2<1.5$ are excluded by LHC. The correction $T_{T^+}$ decreases as $\lambda_2$ increases, and in the allowed regions we find that $T_{W_H} \gg T_{T^+}$. We conclude that the correction from $T$-odd gauge loops to the T-parameter is the dominant constraint in the allowed region where $\lambda_2$ is large. We find the following lower bound on $f$ from \Eq{TWH} at $3\sigma$ after taking $\Lambda = 4 \pi f$
\begin{align}
f
>
(1240~\text{GeV})\times \sqrt{1-\frac16 \log (1+r^2)} \approx (970~\text{GeV}) \times \left(1-0.08(r-3))\right)\,.
\end{align}
Therefore we set the lower bound on the symmetry breaking scale to be $f > 1~\text{TeV}$.
\section{Dark matter phenomenology} \label{dmPheno}
\subsection{Spectrum}
The lightest $T$-odd particle (LTP) in the spectrum is stable and therefore a natural DM candidate. One possible LTP is the gauge field $B_H$. This possibility has been considered in the past in the context of the original LHT model~\cite{Birkedal:2006fz}. In this work we explore the possibility of DM being part of the composite scalar sector, in particular the singlet $s$. The singlet mass $m_s$ is a free parameter in our model. The mass $m_{B_H}$, given in \Eq{gaugeMasses}, is of order $O(200)$~GeV. The region in which $s$ is the LTP corresponds to $r \sim 2-3$ and thus would be the focus of our study. In this region we may safely neglect co-annihilation effects of $s$ with $B_H$.
Since larger values of $r$ correspond to heavier $T$-odd gauge bosons, there is a small increase in the fine-tuning of the Higgs mass from the gauge sector.  We can easily see  by comparing the logarithmically divergent contributions to the Higgs mass from the two sectors
\begin{align}
\frac{\mu^2_{\text{\tiny gauge}}}{\mu^2_{\text{\tiny top}}} \sim \frac{g^4 (1+r^2)}{\lambda_1^2\lambda_2^2} \sim\left(\frac{g^4}{\lambda^2_2} \right)(1+r^2) \sim \text{a few precent}\times (1+r^2)  \,.
\end{align}
that this increase is negligible compared to the dominant source of tuning from the top sector.

\subsection{Singlet-triplet mixing}
The last term of the scalar potential in \Eq{scalarPotentialEq} induces mixing between the singlet $s$ and the neutral component of the triplet $\varphi_3$ after EWSB. The effects of singlet-triplet mixing on the DM phenomenology have been considered in \Ref{Fischer:2013hwa}. 
We focus on the composite nature of the singlet DM. For simplicity, we limit ourselves to the region in parameter space where we may neglect the mixing effects. The mixing angle is given by
\begin{align}
\sin^2\theta_{s\varphi} =\frac{1}{2}\left[1-\sqrt{\frac{1}{1+t^2}}\right] = \frac{t^2}{4}+O(t^4) \,, \;\; t \equiv \frac12\frac{\lambda_\varphi  v^2}{ |m_\varphi^2-m_s^2|}\,.
\end{align}
Assuming for simplicity that $m_s \sim v, \lambda_\varphi \sim 1$ and demanding conservatively  that $\sin\theta_{s\varphi}~<~ 5\%$, we find the following lower bound
\begin{align}
\frac{m_\varphi}{m_s} \gtrsim 2.5\,,
\end{align}
which implies $m_\varphi \gtrsim 600~$GeV.  We note that the assumption $ \lambda_\varphi \sim 1$ as well as the lower bound on $m_\varphi$ are consistent with the IR contribution of \Eq{scalarPotentialEq} to these operators. We find that the operator corresponding to $\lambda_\varphi$ enjoys an accidental factor $\sim5$ enhancement to its coefficient in the CW potential. The IR contributions can be found in \App{scalarPotential} in \Eqs{tripletSingetMixing}{tripletMass}. We conclude that a moderate mass separation is sufficient in order to neglect the singlet-triplet mixing effects.
\subsection{Annihilation cross section} 
The DM relic abundance is calculated by solving the Boltzmann equation for the particle density~\cite{Kolb:1990vq}
\begin{align}
\dot{n}_s +3 H n_{s} = -\left<\sigma v \right> \left[n_s^2-(n_s^{\text{\tiny EQ}})^2\right]\,.
\end{align}
The thermally averaged cross section for a non-relativistic gas at temperature $T$ is given by~\cite{Gondolo:1990dk} 
\begin{align}
\left<\sigma v \right> = \frac{1}{8m_s^4 T K^2_2(m/T)}\int_{4m_s^2}^{\infty} \mathrm{d}s \; \sigma(s-4m_s^2)\sqrt{s} K_1(\sqrt{s}/T) 
\end{align}
and the usual approximation yields~\cite{Kolb:1990vq}
\begin{align}
\Omega_s h^2 \approx 0.12 \left( \frac{3\times 10^{-26} \;\text{cm}^3\; \text{s}^{-1}}{\left<\sigma v \right> }\right) = 0.12 \left(\frac{1\; \text{pb}\; \text{c}}{\left<\sigma v \right> }\right)\,. \label{relicAbundanceAproxx}
\end{align}
The measured DM relic abundance is~\cite{Ade:2013zuv}
\begin{align}
\Omega_{\text{\tiny DM}} h^2 = 0.1199 \pm 0.0027\,.
\end{align}
In the following we consider three types of interactions relevant to our model that determine the annihilation cross section, the Higgs portal, the derivative couplings and the contact term~\cite{Frigerio:2012uc,Marzocca:2014msa}.
%%%%%%%%%%%%%%%%%%%%%%%%%%%%%%%%%%%%%%%%%%%%%%%%%
\subsubsection{Higgs portal}
Due to the explicit breaking of the global symmetry, the scalar potential of \Eq{scalarPotentialEq} is generated radiatively, and in particular the following operators are present in the theory
\begin{align}
\Lagr \ni -\frac12 \tilde m_s^2 s^2 - \ldm s^2 H^\dagger H\,. \label{scalarDMpotential}
\end{align}
$ \ldm $ is the usual Higgs portal coupling of the singlet DM model~\cite{Silveira:1985rk,McDonald:1993ex,Burgess:2000yq}. The Higgs mediates $s$-channel annihilation to SM gauge fields and fermions. The annihilation channel $s s\to h h$ is also possible via the $s$,$t$ and $u$ channels as well as directly via the dimension 4 operator $s^2 h^2$.  We assume that freeze-out occurs after the EW phase transition. In unitary gauge, we can rewrite \Eq{scalarDMpotential} as
\begin{align}
\Lagr \ni -\frac12 \left(\tilde m_s^2+\ldm v^2\right) s^2 - \ldm  v \;  s^2 h- \frac12\ldm s^2 h^2\,.
\end{align}
We define the physical mass of the singlet
\begin{align}
m_s^2 \equiv \tilde{m}_s^2+\ldm v^2\,.
\end{align}
As discussed in \Sec{ScalarPotentialSec}, we take $m_s, \ldm$ to be free parameters. We note a posteriori that the phenomenologically viable regions not excluded by direct detection have $\ldm \lesssim 1\%$. The naive IR contribution to $\ldm$ is $O(10\%)$. To obtain a viable model therefore we assume that additional contributions from UV physics and higher loops, that are expected to be comparable to the leading log ones, generate cancellations of order a few for this coupling to take smaller values.
%%%%%%%%%%%%%%%%%%%%%%%%%%%%%%%%%%%%%%%%%%%%%%%%%
\subsubsection{Goldstone derivative interaction}
The kinetic term of the non linear sigma model of \Eq{nlsigma} contains derivative interactions among the Goldstone fields, in particular
\begin{align}
\Lagr_{nl\sigma }  \ni  \frac{5}{12 f^2}  s_0^2\left[ s (\partial_\mu s)\partial^\mu (H^\dagger H)-  s^2 (\partial^\mu H^\dagger \partial_\mu H)-  (\partial_\mu s )^2 H^\dagger H\right]\,. \label{derInt}
\end{align}
These derivative interactions scales like $m_s^2 /f^2$, and we expect them to become increasingly stronger for heavier DM masses or lower values of $f$. They effect all the annihilation channels of the Higgs portal couplings, typically resulting in destructive interferences.. We discuss this effect in detail in \Sec{relicAbundanceSec}. As $r$ increases, $s_0$ increases and approaches unity.  This is equivalent to decreasing the effective scale of this operator $\tilde f = f / s_0$, thus making these interactions stronger for lower DM masses. 
%%%%%%%%%%%%%%%%%%%%%%%%%%%%%%%%%%%%%%%%%%%%%%%%%
\subsubsection{Contact term}
The non-renormalizable nature of the theory and the mixing in the top sector leads to the appearance of the following contact term, 
\begin{align}
\Ltop \ni \frac{c_{s^2 \bar t t}}{f} s^2 \bar{t}t\,. \label{contactInt}
\end{align}
with
\begin{align}
c_{s^2 \bar t t} =  -y_t s_0^2 \left(\frac{2 \sqrt{2} }{5 }\right) \left(  c_L\left( \frac{7 \sqrt{\xi }}{12 } \right)+s_L\right)\,.
\end{align}
As opposed to the standard singlet DM which interacts with the SM only through the Higgs portal, this dimension 5 operator allows the singlet to annihilate directly into tops without the mediation of the Higgs. Similarly to the derivative interactions, the contact term becomes increasingly important at higher energies. At leading order in $\xi$, we obtain
\begin{align}
c_{s^2 \bar t t}  \approx s_0^2  \left(\frac{ 7 }{15 \sqrt{2} }\right)   \sqrt{\xi }\left(  1 +\frac{12}{7 \lambda_2^2}\right)+O(\xi)\,.
\end{align}
As $r$ increases, the effective scale of this operator $\tilde f = f / \sqrt{s_0}$ decreases, thus making this interaction stronger for lower DM masses. 
%%%%%%%%%%%%%%%%%%%%%%%%%%%%%%%%%%%%%%%%%%%%%%%%%
\subsection{Relic abundance} \label{relicAbundanceSec}
%%%%%%%%%%%%%%%%%%%%%%%%%%%%%%%%%%%%%%%%%%%%%%%%%
We can characterize the DM phenomenology in 3 distinct mass regions, see also~\cite{Bruggisser:2016ixa}. In the first region where $m_s \ll \sqrt{\lambda_{\text{\tiny DM}}}f$, all the effects of the interactions originating from higher dimensional operators, namely the derivative interactions and contact term, are negligible compared to the portal coupling interaction.  The DM phenomenology in this region coincides with the standard singlet DM~\cite{Silveira:1985rk,McDonald:1993ex,Burgess:2000yq}. In regions where $m_s \sim \sqrt{\lambda_{\text{\tiny DM}}}f$, the effect of higher dimensional operators becomes comparable with the marginal portal coupling operator. In particular we find a destructive interference between the Higgs portal coupling and the derivative interactions.  Lastly, for heavy DM masses $m_s \gg \sqrt{\lambda_{\text{\tiny DM}}}f$, the higher dimensional derivative operators dominate. For the following discussion it would be useful to parameterize the thermal cross section as 
\begin{align}
\left< \sigma v \right> =  \sigma_0\left( x\right)\left[\left(\ldm- f_1\left( x\right) \right)^2+f_2(x)\Theta(m_s-m_t)\right]\,,\;\;\; x \equiv \frac{m_s}{f}\,.
\end{align}
$\sigma_0,f_1,f_2$ are monotonically increasing functions of $x$. Furthermore, $\sigma_0,f_1,f_2$ depend in general on $f,r,\lambda_2$. $f_1(x)$ parametrizes the destructive effects of the dimension 6 operator of \Eq{derInt}, hence we expect $f_1 \sim x^2$ .  $f_2(x)$ accounts for the dimension 5 operator of \Eq{contactInt}, which allows the singlet to annihilate into two tops independently of the Higgs interactions, therefore we expect $f_2 \sim x$. 
%%%%%%%%%%%%%%%%%%%%%%%%%%%%%%%%%%%%%%%%%%%%%%%%%
\subsubsection{Portal coupling dominance}
In regions of parameter space where
\begin{align}
m_s \ll \sqrt{\lambda_{\text{\tiny DM}}}f\,,
\end{align}
the composite features of the DM are negligible, and the phenomenology is that of the standard singlet DM~\cite{Silveira:1985rk,McDonald:1993ex,Burgess:2000yq}, where irrelevant operators are irrelevant. In this area of parameter space, the thermally averaged cross section is approximately
\begin{align}
\left< \sigma v \right> \approx  \sigma_0\left( x\right)\ldm^2\,,
\end{align}
and the observed relic abundance is produced for
\begin{align}
\ldm^+(x)  \approx \sqrt{\frac{1~\text{pb}}{\sigma_0\left( x\right)}}\,.
\end{align}
For $\ldm < \ldm^+$ the singlet is over-abundant. These regions are experimentally excluded. In the range $ \ldm > \ldm^+$ the singlet is under-abundant. In this region an additional source of DM must be present in order to account for the observed relic abundance. For a fixed value of $f$, this region is characterized by a large portal couplings or small DM masses. The mass region $m_s < m_h/2$ is severely constrained by the LHC due to the Higgs invisible width to singlets. For $m_s \approx m_h/2$, the Higgs mediator is resonantly produced and $\lambda_{\text{\tiny DM}}$ must be extremely suppressed in order to produce the  correct relic abundance, making this finely tuned region hard to probe experimentally.  We shall focus on DM masses above $m_h/2$ the avoid the above-mentioned issues. 

This region can be seen in the left panel of \Fig{dmpheno} where $m_s<150~$GeV. In this region the total annihilation cross section for a fixed portal coupling decreases with $m_s$, as expected in the standard singlet DM scenario for $m_s > m_h/2$. In the right panel of \Fig{dmpheno}, the portal coupling dominance region is to the right of the minima of the curves. In this region, for a fixed value of the mass, the total annihilation cross section increases with $\ldm$. 
%%%%%%%%%%%%%%%%%%%%%%%%%%%%%%%%%%%%%%%%%%%%%%%%%
\subsubsection{Contact term dominance}
In region of masses where
\begin{align}
 m_s  \sim \sqrt{\lambda_{\text{\tiny DM}}}f\,,
\end{align}
the derivative interactions and Higgs portal are comparable. In this region $\ldm \sim x^2 \sim f_1(x)$ such that the portal coupling and derivative interactions interfere destructively, implying that
\begin{align}
\left< \sigma v \right>  \approx \sigma_0f_2(x)\Theta(m_s-m_t)\,.
\end{align}
In regions where $x < m_t/f$, $\left< \sigma v \right>$ becomes arbitrarily small and the singlet is over-abundant. 
This parameter space is experimentally excluded. In the range where where $x > m_t/f$ we find that $\left< \sigma v \right>$ is positive since the singlet is kinematically allowed to decay into tops. For a particular value $x = x_{\text{\tiny max}}$ defined by
\begin{align}
\sigma_0(x_{\text{\tiny max}})f_2(x_{\text{\tiny max}}) = 1~\text{pb}\,,
\end{align}
the observed relic abundance is produced. In the parameter space where $m_t/f < x < x_{\text{\tiny max}}$ we find that $\left< \sigma v \right> <  1~\text{pb}$ and the singlet is over-abundant. This range is also experimentally excluded. For coupling and masses such that $ x_{\text{\tiny max}}<x$ we find that $\left< \sigma v \right> >  1~\text{pb}$ and the singlet is under-abundant. In this region an additional source of DM must be present in order to account for the observed relic abundance. We conclude that for a given point in $(\lambda_2,r,f)$ parameter space, the largest DM mass for which the singlet can account for the entire DM relic abundance is therefore given by $m_s^{\text{\tiny max}} =  \sqrt{x_{\text{max}}}f$.

The relevant parameter space in the left panel of \Fig{dmpheno} corresponds to the region where $m_s \sim 220$~GeV, close to the minimal value of the cross section. The annihilation to the Higgs and gauge bosons is effectively suppressed by the destructive interference between the portal coupling and the derivative interactions. As this suppression occurs where $m_s>m_t$, the remaining annihilation cross section is exclusively to tops. In the right panel of \Fig{dmpheno}, the minima of the different curves are precisely mapped to this area of maximal interference. For the fixed mass $m_s=150$~GeV, the singlet is not allowed kinematically to decay into tops and the annihilation cross section vanishes. Conversely, for $m_s=200$~GeV the decay into tops is allowed and the annihilation cross section is dominated by the contact term. Lastly, the minimum of the curve corresponding to $m_s=250$~GeV is approximately $1$~pb, meaning that for this particular point in the $(\lambda_2,r,f)$ parameter space, $x_{\text{\tiny max}} \approx 250/1000 = 1/4$.
%%%%%%%%%%%%%%%%%%%%%%%%%%%%%%%%%%%%%%%%%%%%%%%%%
\subsubsection{Derivative interaction dominance }
\begin{figure}[h!]
\center
\includegraphics[width=0.49\textwidth]{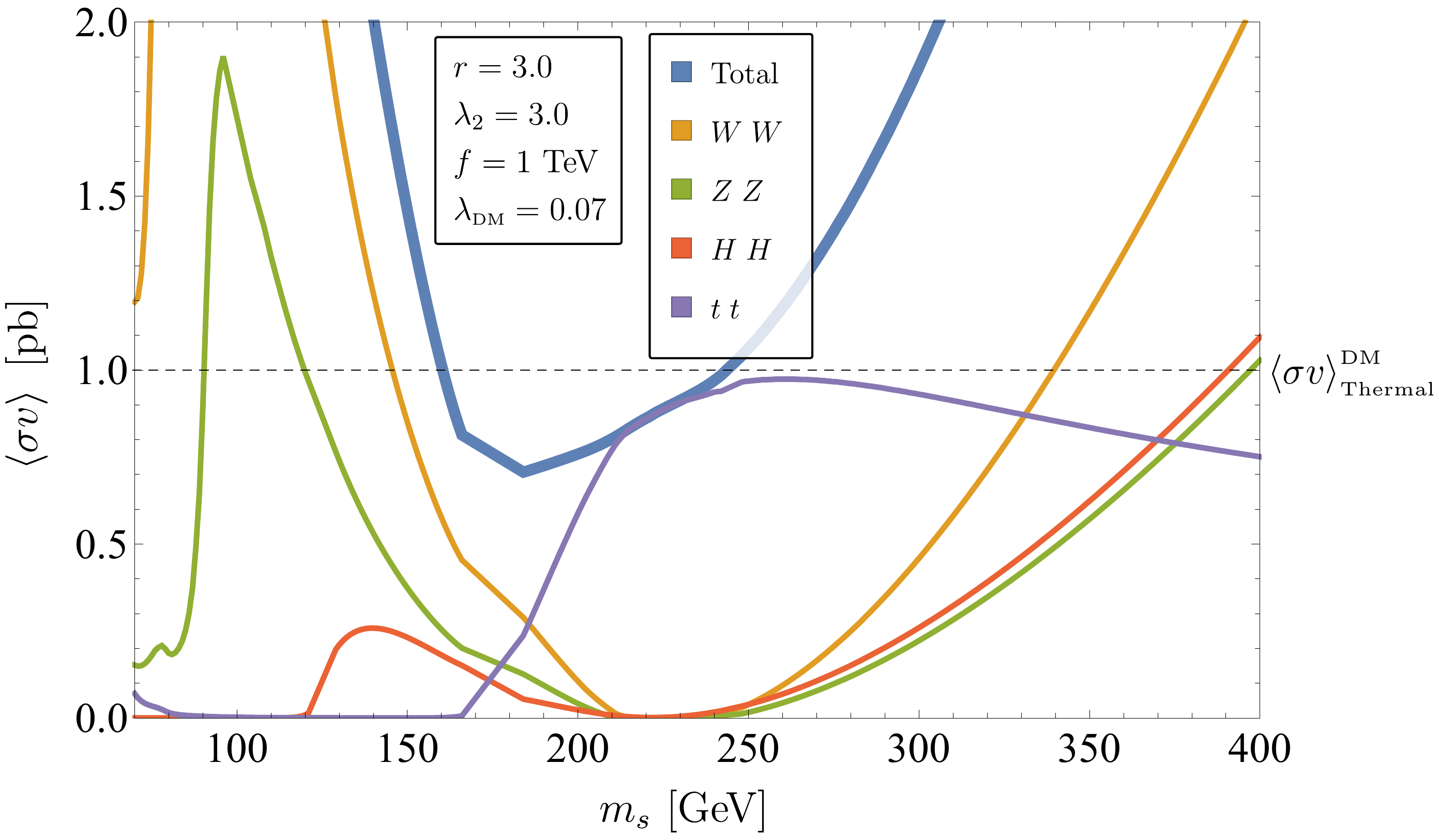}
\includegraphics[width=0.49\textwidth]{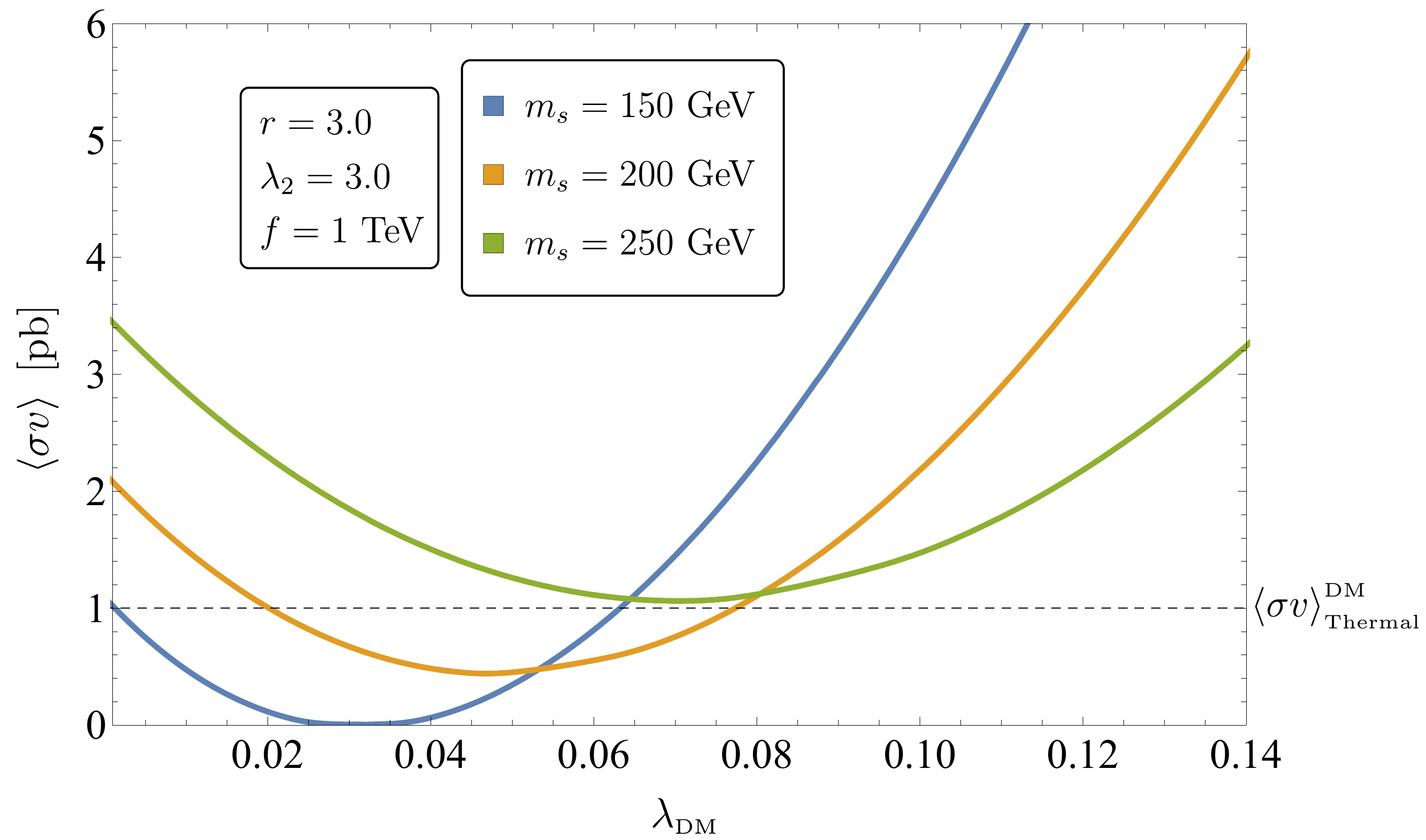}
\caption{\emph{Left panel:} The thermally averaged cross section as a function of the DM mass $m_s$ for $\ldm~=~0.07,\,f~=~1000~\text{GeV},\,r=3$ and $\lambda_2=3$. 
The dashed line at $\left< \sigma v \right> = 1$~pb  represents the cross section that produces the correct relic abundance according to \Eq{relicAbundanceAproxx}. \emph{Right panel:} The thermally averaged cross section as a function of $\ldm$ for different values of $m_s$ with $f=1000~\text{GeV}\,,r=3$ and $\lambda_2=3$. The dashed line at $\left< \sigma v \right> = 1$~pb represents the cross section that produces the correct relic abundance according to \Eq{relicAbundanceAproxx}. 
}
\label{dmpheno}
\end{figure}

In the regions of parameters space where
\begin{align}
m_s  \gg \sqrt{\lambda_{\text{\tiny DM}}}f\,,
\end{align}
the irrelevant operators, namely the dimension 6 operators corresponding to the derivative interactions, are dominating, and the annihilation cross section grows with the singlet mass. The observed relic abundance is produced for
\begin{align}
\ldm^{-} \approx f_1(x)-\sqrt{\frac{1 \text{ pb }}{ \sigma_0\left( x\right)}-f_2(x)\Theta(m_s-m_t)}\, \;\;\text{for} \;\;x>x_{\text{\tiny min}}\,,
\end{align}
with $x_{\text{min}}$ defined by 
\begin{align}
f_1(x_{\text{min}})=\sqrt{\frac{1 \text{ pb }}{ \sigma_0\left( x_{\text{min}}\right)}-f_2(x_{\text{min}})\Theta(x_{\text{min}}-m_t/f)}\,.
\end{align}
For $x \sim x_{\text{min}}$ the correct relic abundance is recovered with $\ldm^- \ll 1$ and with DM mass $m_s^{\text{\tiny min}} \equiv  \sqrt{x_{\text{\tiny min}} }f$. The nuclear cross section is typically $\sim10^{-11}$~pb, beyond the reach of current direct detection experiments. For $\ldm > \ldm^-$ the singlet is over-abundant. This regions are experimentally excluded. In the region $ \ldm < \ldm^-$ the singlet is under-abundant. In this region an additional source of DM must be present in order to account for the observed relic abundance.
\\ \\
In the left panel of \Fig{dmpheno}, the derivative interactions become dominant at $m_s>225$~GeV. The total annihilation cross section increases with $m_s$ for a fixed $\ldm$, and the annihilation channels to the Higgs and gauge bosons become dominant compared to the annihilation channel to tops. In the right panel of \Fig{dmpheno} the derivative interactions dominance region can be identified to the left of the minima, where $\ldm$ is small. The annihilation cross section increases as $\ldm$ decreases. In this region smaller values of $\ldm$ correspond to smaller destructive interference between the portal coupling and the derivative interactions, and therefore an increased overall annihilation cross section. For the curve corresponding to $m_s=150$~GeV, we see that $\ldm^+ \approx 0.065$ and $\ldm^- \ll 1$, meaning that for this particular point in the $(\lambda_2,r,f)$ parameter space, $x_{\text{\tiny min}} \approx 150/1000 = 0.15$. 
\subsection{Direct detection}
\begin{figure}[h!]
\center
\includegraphics[width=0.32\textwidth]{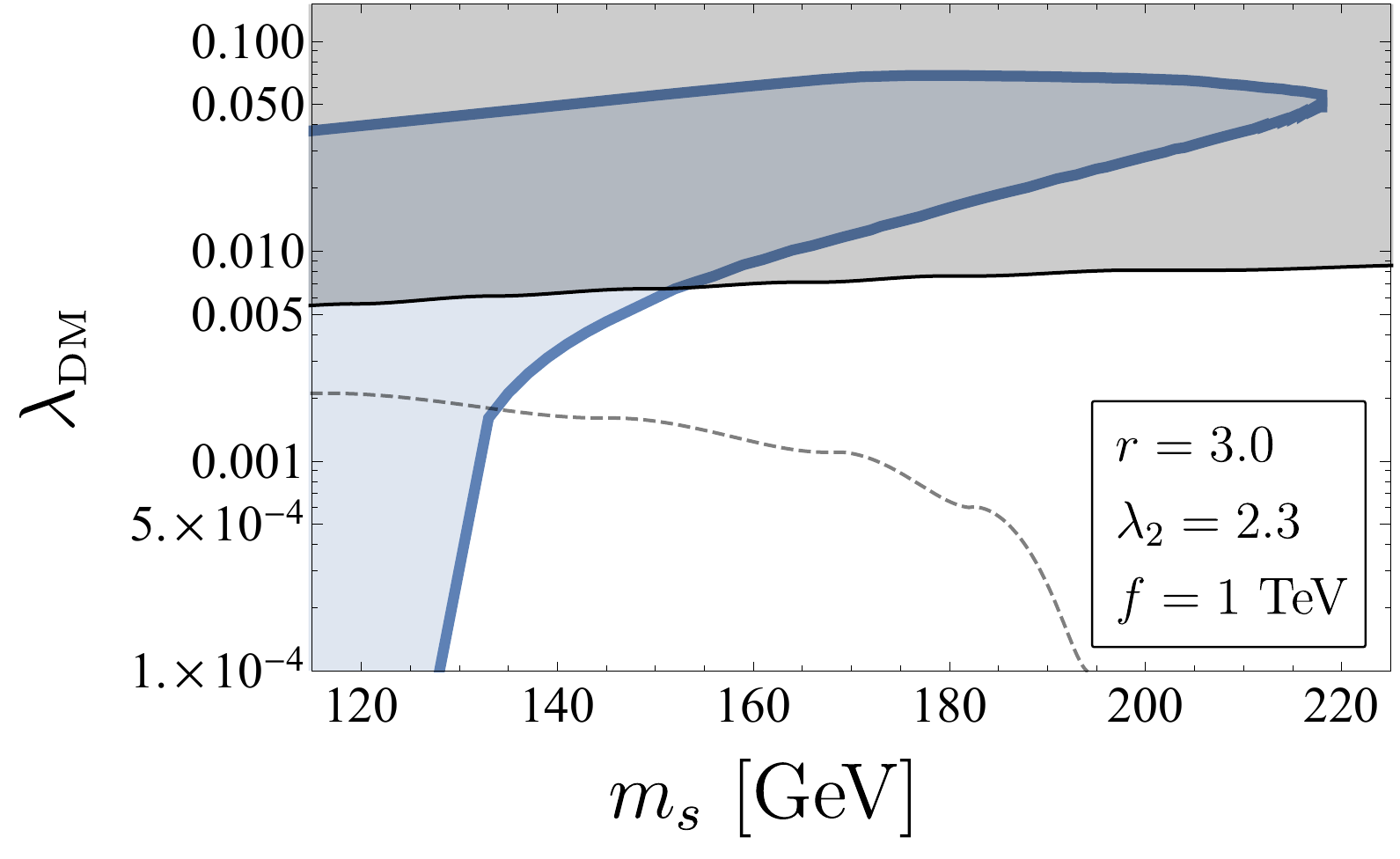}
\includegraphics[width=0.32\textwidth]{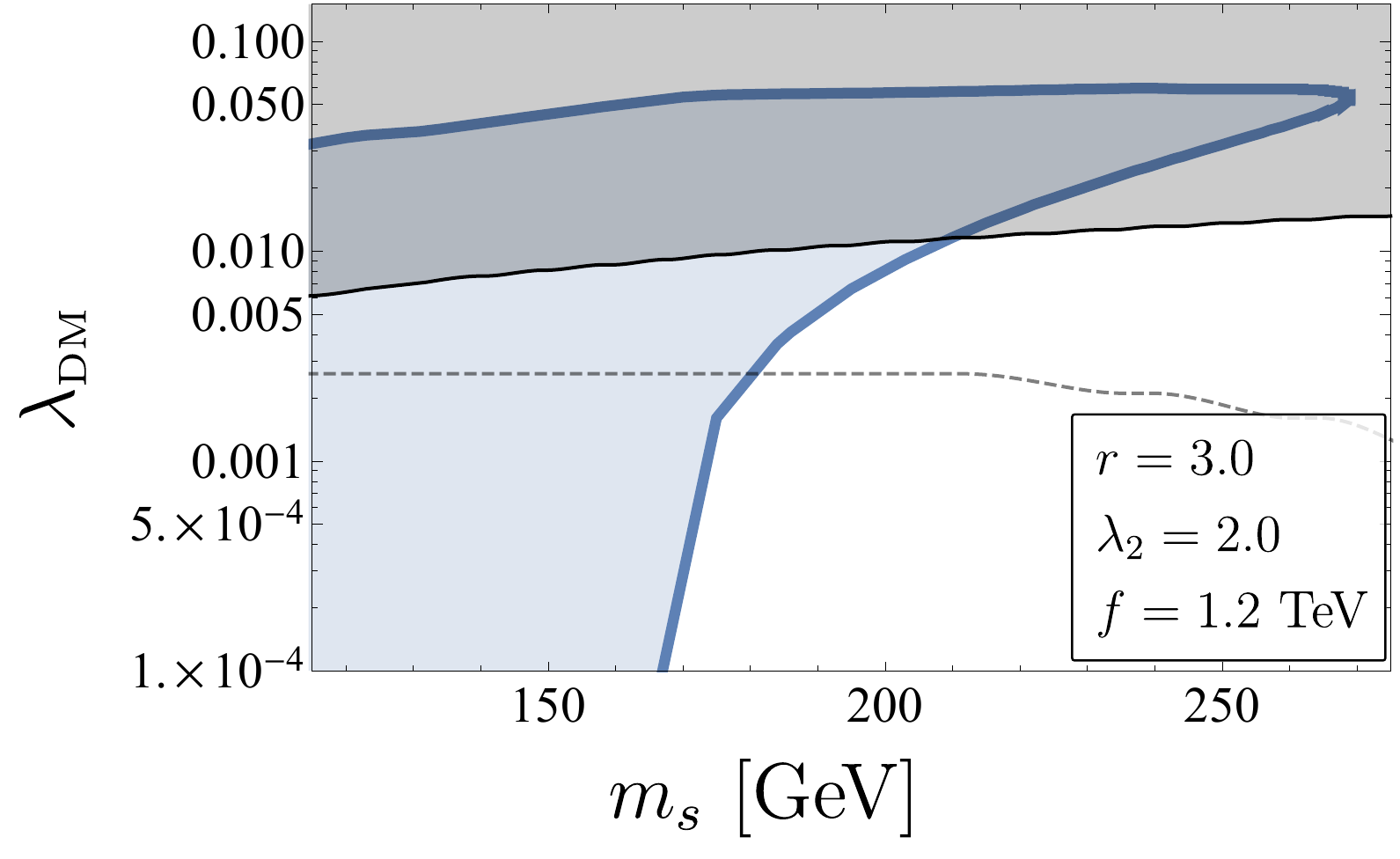}
\includegraphics[width=0.32\textwidth]{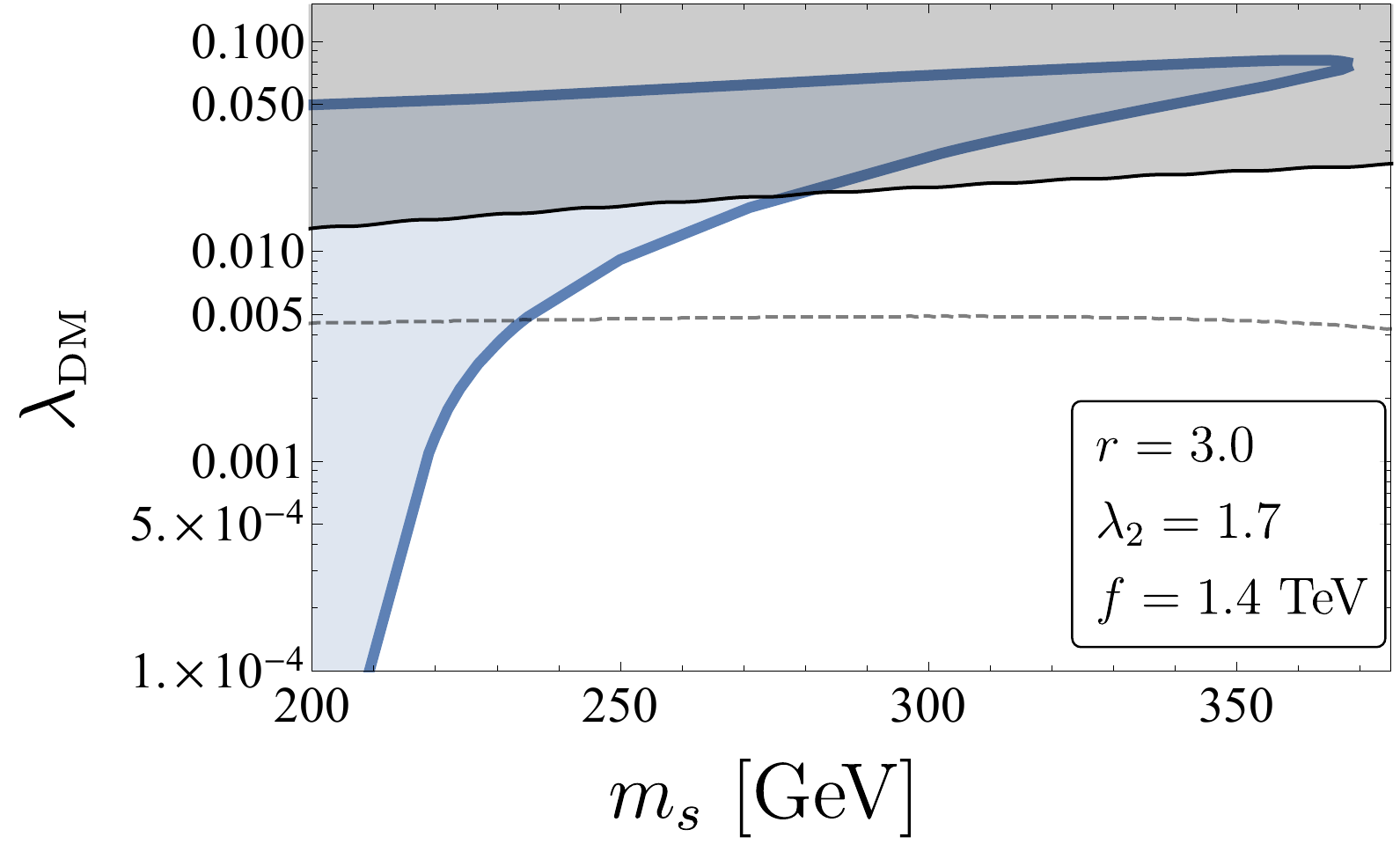}
\caption{Singlet relic abundance in the $m_s,\ldm$ plane for $f=1$~TeV (left), $f=1.2$~TeV (middle) and $f=1.4$~TeV (right), for fixed $r=3$ and minimal $\lambda_2 \sim \frac{2300~\text{GeV}}{f}$. The solid blue lines represent areas where $\Omega_s  = \Omega_{\text{\tiny DM}} $. The blue areas are regions where $\Omega_s  > \Omega_{\text{\tiny DM}} $, and therefore are excluded. The grey regions are excluded by XENON1T~\cite{Aprile:2017iyp} after 34.2 live days. The Dashed lines are the projected sensitivities for XENON1T at $1.1~\text{yrs $\times$ Ton}$~\cite{Aprile:2015uzo}.  
}
\label{dmrelic}
\end{figure}

\begin{figure}[h!]
\center
\includegraphics[width=0.65\textwidth]{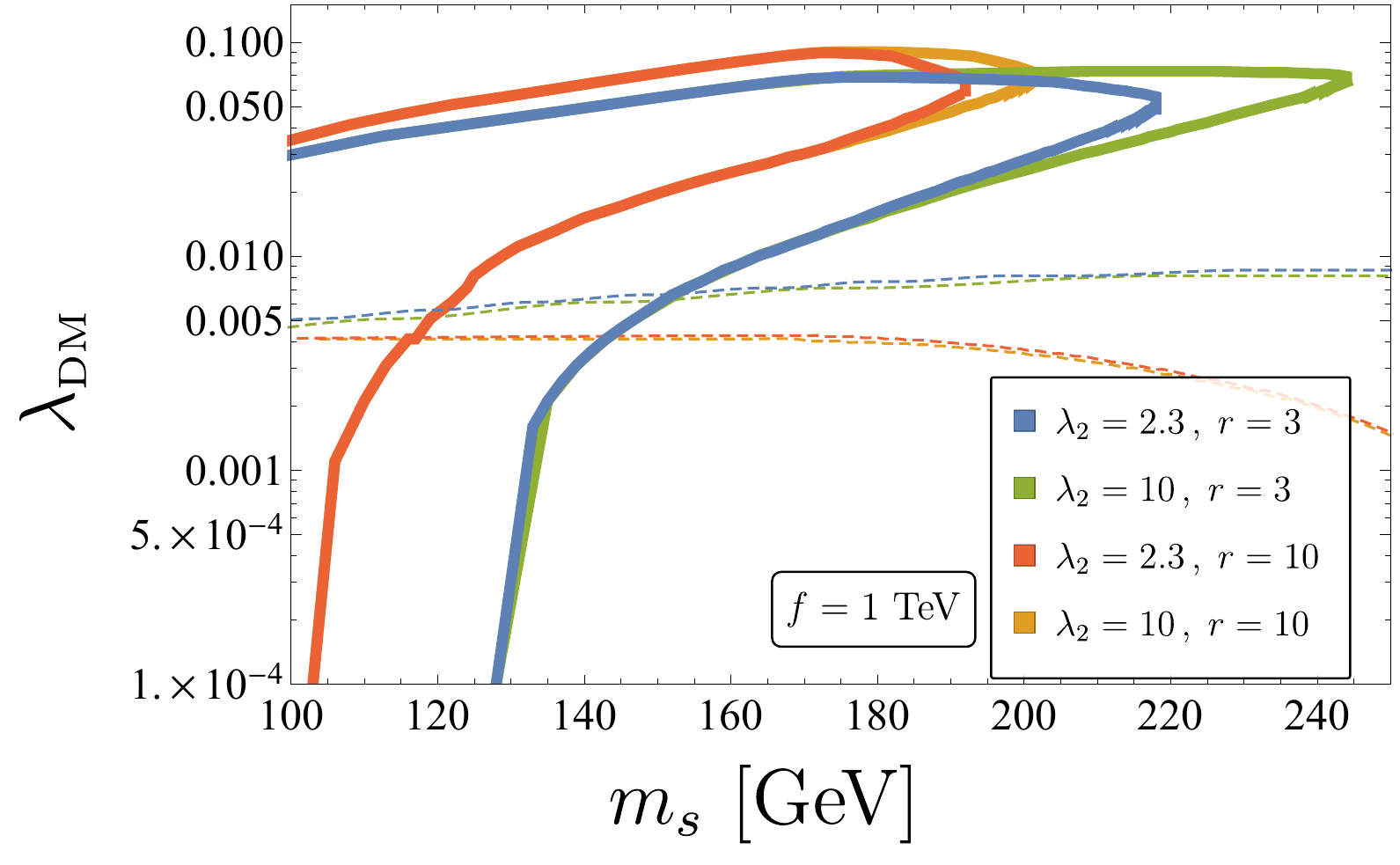}
\caption{The effects of changing $r$ and $\lambda_2$ on the relic abundance curves, shown as solid curves. The dashed curves represent the XENON1T~\cite{Aprile:2017iyp} bounds after 34.2 live days. Increasing $r$ has similar effects to lowering $f$ - the coefficients of the non-renormalizable terms increase and their effect is noticeable at lower DM masses. Increasing $\lambda_2$ reduces the size of the coefficient of the dimension 5 contact term, therefore increasing $m_s^{\text{\tiny max}} =  \sqrt{x_{\text{max}}}f$.   }
\label{dmrelicDiffRlam}
\end{figure}

The model was implemented using {\sc FeynRules}~\cite{Alloul:2013bka} and exported to \texttt{micrOMEGAs}~\cite{Belanger:2014vza}. The strongest direct detection bounds are due to XENON1T~\cite{Aprile:2017iyp} after 34.2 live days. Scan results for this model can be seen in \Fig{dmrelic}. The two branches appearing in each panel represent the two possible solutions for $\ldm$ for each mass value which produce the observed relic abundance. The branches meet at some maximal DM mass, above which the singlet is always under-abundant. The upper branch is ruled out by direct detection. Some of the lower branch is still consistent with experimental bounds.  In the region where $m_s \approx \sqrt{x_{\text{\tiny min}}}f$, $\ldm$ can be arbitrarily small, thus avoiding direct detection. In this regions, the theory gives a sharp prediction for the DM mass. At mentioned previously, the naive IR contribution to $\ldm$ is too big and of $O(10\%)$. We therefore assume that additional contributions from UV physics and higher loops generate mild cancellations, allowing this coupling to take the allowed $O(1\%)$ values. 
\\ \\The impact of varying $\lambda_2,r$ for a fixed value of $f$ can be seen in \Fig{dmrelicDiffRlam}. The largest effect is seen for increasing $r$, which in turn raises the importance of the non-renormalizable interactions at lower DM masses. A smaller effect due to the increase of $\lambda_2$ can be seen in the meeting point of the two branches. Larger values of $\lambda_2$ decrease the contact term, pushing $m_s^{\text{\tiny max}} =  \sqrt{x_{\text{max}}}f$ to higher values.
\newpage
\section{Conclusions} \label{conc}
In this work we have presented a viable composite dark matter (DM) candidate within the Littlest Higgs with $T$-parity framework. We started by motivating a minimal extension of the original coset which allows the $T$-odd doublet to acquire a mass without introducing additional sources of explicit symmetry breaking.  The extended coset contains a $T$-odd electroweak singlet. This singlet is naturally light and therefore it is reasonable to assume it is the lightest $T$-odd particle, which insures its stability. 
\\ \\
The top sector is implemented using a collective breaking mechanism, insuring the absence of quadratically divergent contribution to the Higgs mass at 1 loop. $T$-parity implies a rich LHC phenomenology:  in addition to the usual ($T$-even) top partners, the top sector contains $T$-odd top partners. This $T$-odd top partners can be doubly produced via QCD in the LHC and decay to standard model (SM) particles and missing energy. We have derived lower bounds on the masses of the $T$-even and $T$-odd top partners from various LHC searches. When combined with electroweak-precision-test (EWPT) bounds, we derived a set of constraints on the parameter space of the model. 
\\ \\ 
We examined the DM phenomenology of the composite singlet DM within the allowed parameter space. The usual "elementary" singlet DM scenario is heavily constrained by direct detection experiments. In the composite singlet DM scenario, the composite nature of the DM allows it to escape detection in areas with $O(1\%)$ portal coupling, while still producing the observed relic abundance via its derivative interactions with the Higgs. 
The "elementary" singlet can only hide in the finely tuned "resonance" valley where $m_s \approx m_h/2$. Conversely, the composite singlet can exist in a broader region, corresponding to different values of $f$ and $r$, in which it can evade detection. In these regions the correct relic abundance can be produced only due to the derivative interactions. The small portal coupling needed in these regions would in general require some mild amount of fine tuning, unless one can find a way to suppress it e.g using symmetries or additional dynamics.
\acknowledgments
We thank D. Pappadopulo, M. Ruhdorfer, E. Salvioni, and A. Vicchi for useful discussions. The work of RB is supported by the Minerva foundation. The work of GP is supported by grants from the BSF, ERC, ISF, Minerva, and the Weizmann-UK Making Connections Programme. RB and AW have been partially supported by the DFG cluster of excellence EXC 153 "Origin and Structure of the Universe", by the Collaborative Research Center SFB1258, the COST Action CA15108, and the European Union's Horizon 2020 research and innovation programme under the Marie Curie grant agreement, contract No. 675440.
\appendix
\section{The complete Lagrangian} \label{modelApp}
The model is defined by the global symmetry
\begin{align}
\mathcal{G} = SU(5) \times SU(2)_L \times SU(2)_R \times U(1)_L \times U(1)_R \times U(1)_Q\,. \label{globalGr}
\end{align}
A global unbroken $U(1)_Q$ is added in order to fix the hyper charges of the matter fields. $SU(5)$ contains two $ SU(2)\times U(1)$ subgroups defined in \Eqs{su2l}{su2r}, denoted by $[SU(2)\times U(1)]_{1/2}$.  We gauge the following subgroup
\begin{align}
SU(2)_{1+L}\times U(1)_{1+L+Q} \times SU(2)_{2+R}   \times U(1)_{2+R+Q}\,.
\end{align}
We implicitly include $SU(3)_c$ as an external gauge symmetry. We introduce fields in representations of $\mathcal{G}$ denoted by $ (\mathcal{R},\mathcal{R}_L,\mathcal{R}_R)_{q_L,q_R,q_Q}$. A generic representation of \Eq{globalGr} is mapped under $T$-parity to
\begin{align}
\text{$T$-parity}:\,(\mathcal{R},\mathcal{R}_L,\mathcal{R}_R)_{q_L,q_R,q_Q} \to (\overline{\mathcal{R}},\mathcal{R}_R,\mathcal{R}_L)_{q_R,q_L,q_Q}\,.
\end{align}
The representation $\overline{\mathcal{R}}$ is defined by the automorphism of \Eq{Tparitydef}. The Lagrangian is described by the following sum
\begin{align}
\Lagr = \Lagr_{\text{gauge}}  + \Lagr_{\text{kin}} + \Ltop + \Lagr_{\kappa}\,.
\end{align}
The gauge kinetic terms are given as usual by
\begin{align}
\Lagr_{\text{gauge}} = -\frac14 \sum_{i=1,2}W_{ia}^{\mu\nu} W^{ia}_{\mu\nu} -\frac14 \sum_{i=1,2}B_i^{\mu\nu} B^i_{\mu\nu} -\frac14 G_a^{\mu\nu} G^a_{\mu\nu}\,.
\end{align}
We introduce two scalar fields with the following $\mathcal{G}$  representation 
\begin{align}
\Sigma : ( {\bf{15}} ,{\bf{1}},{\bf{1}})_{0,0,0}\,, \;\; X : ( {\bf{1}} ,{\bf{2}},\overline{{\bf{2}}})_{q_X,-q_X,0}\,.
\end{align} 
The charges of $X$ under $U(1)_L \times U(1)_R$ are constrained by the requirement to preserve the $T$-even combination $U(1)_{R+L}$.  We determine the value of $q_X$ in \Eq{qXqpsi}. The global symmetry is spontaneously broken by the VEV's of $\Sigma$ and $X$
\begin{align}
\frac{SU(5) }{SO(5) }\times \frac{ [ SU(2) \times U(1) ]_{L}\times [ SU(2) \times U(1) ]_{R} \times U(1)_Q} { [ SU(2) \times U(1) ]_{L+R}\times U(1)_Q} \,.
\end{align}
We parametrize $\Sigma$ and $X$ following \Eqs{pi_sigma matrix}{pi_X matrix} and write down the kinetic terms of the non-linear sigma model 
\begin{align}
\Lagr_{\text{nl}\sigma} =  \frac{f^2}{8}\text{Tr}[(D_\mu \Sigma) (D^\mu \Sigma^*)] +  \frac{{f'}^2}{4}\text{Tr}[(D_\mu X) (D^\mu X^\dagger) ] \,,
\end{align}
with
\begin{align}
& D \Sigma = \partial \Sigma -i \sum_{i=1,2} g_i W^a_i ( Q_i^a \Sigma +  \Sigma {Q_i^a}^T)-i  \sum_{i=1,2} g'_i B_i ( Y_i \Sigma +  \Sigma {Y_i}^T)\,,
\\
& D X = \partial X - \frac i2 (g_1 W_1^a \sigma^a X - g_2 W_2^a  X\sigma^a) - i   q_X (g'_1 B_1-g'_2 B_2)\,. \label{XcovDev}
\end{align}
$T$-parity dictates that
\begin{align}
g_1=g_2=\sqrt{2}g\,, \;\;\; g'_1=g'_2=\sqrt{2}g'\,,
\end{align}
with $g,g'$ the SM gauge couplings. 
\\The matter sector contains the following linearly transforming fields
\begin{align}
\Psi_1 = \begin{pmatrix}
\psi_1 \\ \chi_1 \\0
\end{pmatrix} : (\overline{\bf{5}},{\bf{1}},{\bf{1}})_{0,0,\frac13}\,,
\;\;
\Psi_2 = \begin{pmatrix}
0 \\ \chi_2 \\ \psi_2
\end{pmatrix} :  ({\bf{5}},{\bf{1}},{\bf{1}})_{0,0,\frac13}\,,
\end{align}
and 
\begin{align}
\tau_1 :  ({\bf{1}},{\bf{1}},{\bf{1}})_{\frac{8}{15},\frac{2}{15},0} \,,\;\; \tau_2 :  ({\bf{1}},{\bf{1}},{\bf{1}})_{\frac{2}{15},\frac{8}{15},0} \,, \;\;\; \tilde{t}_R  : ({\bf{1}},{\bf{1}},{\bf{1}})_{0,0,\frac{1}{3}}\,.
\end{align}
We introduce a non linearly transforming doublet $\psi^-_R$. Non-linear representations are described in terms of representations of the unbroken subgroup
\begin{align}
\mathcal{H} = SO(5) \times  SU(2)_{L+R}   \times U(1)_{L+R}\times U(1)_Q\,.
\end{align}
 $\psi^-_R$ transforms non-linearly under the full global group $\mathcal{G}$ using the CCWZ formalism. In our case
\begin{align}
\psi^-_R : ({\bf{1}},{\bf{2}})_{q_X,q_\psi} \;\; \text{under} \;\; \mathcal{H}\,.
\end{align}
The $U(1)_Q$ charge of $\psi^-_R$ , denoted here by $q_{ \psi}$, is determined in \Eq{qXqpsi}. $q_X$ is the same charge appearing in \Eq{XcovDev}. Under $T$-parity, 
\begin{align}
\Psi_1 \to \Omega \Sigma_0 \Psi_2\,,\;\;  \tau_1 \leftrightarrow \tau_2\,, \;\;  \tilde{t}_R \to \tilde{t}_R\,,\;\;\psi^-_R \to -\psi^-_R \,.
\end{align}
The $U(1)$ charge assignments are fixed by matching the required SM hyper charges and requiring that all the gauged $U(1)$ symmetries are conserved. The SM hyper charge is given by 
\begin{align}
Y_{\SM}  = Y_1+Y_2+q_L+q_R+2q_Q\,.
\end{align}
e.g for $\psi_1$, 
\begin{align}
Y_{\SM} = (-3/10)+(-2/10)+(0)+(0)+2(1/3) = 1/6\,.
\end{align} 
Let us determine the $q_X$ and $q_\psi$ charges. Defining $U \equiv e^{\frac{i \Pi_X}{f'}}$, the combinations $U \psi^-_R$ and $U^\dagger \psi^-_R$ transform linearly under the global group
\begin{align}
U \psi^-_R :  ({\bf{1}},{\bf{2}},{\bf{1}})_{q_X,0,q_{\psi}}\,, \;\; U^\dagger \psi^-_R :  ({\bf{1}},{\bf{1}},{\bf{2}})_{0,q_X,q_{\psi}}\,.
\end{align}
Using these identifications as linear representations, it is clear that conservation of $U(1)_{1+L+Q}$ and $U(1)_{2+R+Q}$, e.g in the first term of \Eq{Lkappa}, requires
\begin{align}
- \left(-\frac{3}{10}+\frac 13 \right) + q_X + q_\psi = 0 \;\;\;  \text{and} \;\;\;  - \left(-\frac{1}{5}+\frac 13 \right) + q_\psi = 0 \;\;\;\to\;\;\; q_\psi = \frac{2}{15}\,, \;\; q_X = -\frac{1}{10}\, . \label{qXqpsi}
\end{align}
We introduce the kinetic terms
\begin{align}
\Lagr_{\text{\tiny kin}} = i \sum_{i=1,2}\overline{\Psi}_i \slashed{D} \Psi_i +i \sum_{i=1,2}\overline{\tau}_i \slashed{D} \tau_i+ i \overline{\tilde t_R}  \slashed{D} \tilde{t}_R + i \overline{\psi^-_R}\slashed{D} \psi^-_R\,. \label{kinTerms}
\end{align}
The kinetic term for the non-linearly transforming doublet $\psi^-_R$
\begin{align}
D_\mu \psi^-_R = (\partial_\mu + e_\mu -iq_{ \psi} (g'_1B_{1\mu}+ g'_2B_{2\mu})) \psi^-_R\,. \label{qpsi}
\end{align}
The $e_\mu \equiv  e^i_\mu T^i$ symbol of the CCWZ formalism connects the non-linearly transforming field and the NGB's via the matrix $U$\cite{Coleman:1969sm,Callan:1969sn}
\begin{align}
  U^\dagger (D_\mu U) \equiv d^j_\mu X^j + e^i_\mu T^i\,, \;\; D_\mu U  = \left(\partial_\mu -ig_1 W^a_1 \frac{\sigma^a}{2}-i q_X g'_1 B_1\right)\,.
\end{align}
Using the automorphism defined by $T$-parity we can also write
\begin{align}
 U (D_\mu U^\dagger)  \equiv -d^j_\mu X^j + e^i_\mu T^i \,, \;\; D_\mu U^\dagger  = \left(\partial_\mu -ig_2 W^a_2 \frac{\sigma^a}{2}-i q_X g'_2 B_2\right)\,.
\end{align}
This automorphism allows us to write the $e_\mu$ symbol in terms of the pion matrix $U$ and the gauge fields
\begin{align}
 e_\mu = \frac12 \left(   U^\dagger D_\mu U+U D_\mu U^\dagger  \right) \,.
\end{align}
The covariant derivatives of $\Psi_1$ and $\Psi_2$ are
\begin{align}
D_\mu \Psi_1 &= (\partial_\mu+ i \sum_{i=1,2}[g_i W_{i\mu}^a Q^{a*}_i+g'_i B_{i\mu} Y^*_i]- \frac{i}{3}  (g'_1B_{1\mu}+ g'_2B_{2\mu})  )  \Psi_1\,,
\\ D_\mu \Psi_2 &= (\partial_\mu-i \sum_{i=1,2}[g_i W_{i\mu}^a Q^a_i+g'_i B_{i\mu} Y_i]- \frac{i}{3}  (g'_1B_{1\mu}+ g'_2B_{2\mu}))  \Psi_2\,.
\end{align}
The covariant derivative of a singlet field  $\chi$ transforming as $({\bf{1}},{\bf{1}},{\bf{1}})_{q_L,q_R,q_Q}$ is given by
\begin{align}
D_\mu \chi = (\partial_\mu- i q_L  g'_1B_{1\mu} - i q_R   g'_2B_{2\mu}  - i q_Q  (g'_1B_{1\mu}+ g'_2B_{2\mu})  ) \chi\,.
\end{align}
For completeness we report the top sector Lagrangian
\begin{align}
\Ltop &= \frac{\lambda_1 f}{2}\left(  {\overline \Psi_1}_i O_i +   (\overline \Psi_2  \Omega \Sigma_0)_i \tilde O_i \right) \tilde t_R+\frac{\lambda_2 f}{\sqrt{2}}\left( \overline{\chi}_1 \tau_1 - \overline{\chi}_2 \tau_2 \right)+\text{h.c}\,, \nonumber
\\O_i &\equiv  \epsilon_{ijk}\Sigma_{j4}\Sigma_{k5}\,,\;\; \tilde{O}_i \equiv 2 \epsilon_{ijk}\tilde\Sigma_{j4}\tilde\Sigma_{k5}\,,
\end{align}
and the terms that gives the $T$-odd doublet combination a mass
\begin{align}
\Lagr_{\kappa} =  \frac{\kappa f}{\sqrt{2}}\left( \overline \psi_1 \sigma_2 U - \overline \psi_2 \sigma_2 U^\dagger  \right ) \psi_R^- +\hc\,. 
\end{align}
\section{The scalar potential and its symmetries} 
\label{scalarPotential}
In this appendix we discuss in detail the symmetry structure of the model and the scalar potential. The Higgs doublet is protected by two different shift symmetries. Each of the shift symmetries is contained inside a different $SU(3)$ subgroup of $SU(5)$
\begin{align}
 \exp\left[ \frac{i}{\sqrt{2} f}\begin{pmatrix}
&\vec{\epsilon}&\phantom{10}\\
\vec{\epsilon}^T&&\\
&&
\end{pmatrix}\right] \in [SU(3)]_1 \,, \;\; 
 \exp\left[ \frac{i}{\sqrt{2} f}\begin{pmatrix}
\phantom{10}&&\\
&&\vec{\epsilon}^T\\
&\vec{\epsilon}&
\end{pmatrix} \right]\in [SU(3)]_2 \,.
\end{align}
All the couplings that explicitly break the global symmetry in this model, namely the gauge couplings and the top sector couplings, preserve at least one of the $SU(3)$ subgroups. A Higgs potential is generated only when at least two couplings are non zero, such that all the shift symmetries are broken. This so-called "Collective Breaking" mechanism insures the absence of quadratically divergent contributions to the Higgs mass. The couplings and their $T$-parity conjugate respect different symmetries, therefore it is useful to denote the T-conjugate couplings with a tilde 
\begin{align}
\Ltop&=\frac{ f}{4}\left( \lambda_1 {\overline \Psi_1}_i O_i+  \tilde{\lambda}_1 (\overline \Psi_2  \Omega \Sigma_0)_i \tilde O_i \right) \tilde t_R+\frac{ f}{\sqrt{2}}\left( \lambda_2\overline{\chi}_1 \tau_1 - \tilde\lambda_2\overline{\chi}_2 \tau_2 \right)+\hc\,,
\\
\Lagr_{\kappa}  &= \frac{ f}{\sqrt{2}}\left( \kappa\overline \psi_1 \sigma_2 U -\tilde\kappa \overline \psi_2 \sigma_2 U^\dagger  \right ) \psi_R^- +\hc\,.
\end{align}
In order to better understand the structure of the generated scalar potential, we assign spurionic $U(1)$ charges to our fields and couplings, which can be found in \Tab{spurionCharges}.
\begin{table}
\begin{center}
  \begin{tabular}{ |c | c | c| c |c|c| c | c | c| c |c|c| }
    \hline
    $\Psi_1$ & $\Psi_2$ & $t$ &$\tau_1$& $\tau_2$ & $\psi_{R}^-$&  $\lambda_1$ & $\tilde\lambda_1$ & $\lambda_2$ &$\tilde\lambda_2$& $\kappa$ & $\tilde \kappa$ \\ \hline 
    $a$ & $b$ & $c$ & $d$ & $e$ & $f$ &    $a-c$ & $b-c$ & $a-d$ & $b-e$ & $a-f$ & $b-f$\\ \hline
  \end{tabular}
\end{center}
\caption{Spurionic $U(1)$ assignment for the couplings and fields.}
\label{spurionCharges}
\end{table}
The combinations of couplings appearing in the quadratically divergent contribution to the scalar potential must be of the form $g g^\dagger$ or $\tilde g \tilde g^\dagger$. We can deduce from the residual symmetries a generic form for the quadratically divergent potential. For concreteness let us consider the coupling $\lambda_1$ and set all the other explicit symmetry breaking couplings to zero. The original coset 
\begin{align}
\frac{SU(5) }{SO(5) }\times \frac{ [ SU(2) \times U(1) ]_{L}\times [ SU(2) \times U(1) ]_{R} \times U(1)_Q} { [ SU(2) \times U(1) ]_{L+R}\times U(1)_Q} \,,
\end{align}
contains $(24-10)+ (9-5) = 18$ NGB's, out of which 4 are eaten, leaving us with 14 physical NGB's with the following $SU_L(2)\times U_Y(1)$ representations
\begin{align}
\bf{3}_{\pm 1} \oplus \bf{2}_{\pm 1/2} \oplus \bf{3}_0 \oplus \bf{1}_0\,.
\end{align}
Turning on only $\lambda_1$ breaks the global symmetry and changes the coset structure
\begin{align}
\frac{SU(3)\times   [ SU(2) \times U(1) ]_{2}}{ [ SU(2) \times U(1) ]_{1+2}}\times \frac{ [ SU(2) \times U(1) ]_{L}\times [ SU(2) \times U(1) ]_{R} \times U(1)_Q} { [ SU(2) \times U(1) ]_{L+R}\times U(1)_Q} \,,
\end{align}
This coset contains $(8+4)-4 + (9-5) = 12$ NGB's, out of which 4 are eaten, leaving us with 8 physical NGB's with the following $SU_L(2)\times U_Y(1)$ representations
\begin{align}
\bf{2}_{\pm 1/2} \oplus \bf{3}_0 \oplus \bf{1}_0\,.
\end{align}
There must exist a non-linear combination of the goldstone fields 
\begin{align}
\tilde \Phi_{ij} \equiv  f_1\left(1,\frac{s}{f},\frac{s^2}{f^2},\frac{\varphi^2}{f^2},\frac{|H|^2}{f^2},...\right) \Phi_{ij} + f_2\left(1,\frac{s}{f},\frac{s^2}{f^2},\frac{\varphi^2}{f^2},\frac{|H|^2}{f^2},...\right) \frac{H_i H_j }{f}
\end{align}
with $f_1,f_2$ some functions of gauge-invariants, such that the quadratically divergent potential can be written as gauge-invariant function of only $\tilde \Phi$
\begin{align}
V(\Phi,H,s,\varphi) = V(\tilde \Phi)  \label{tildePhiEq}\,.
\end{align}
This constraint limits the form of the quadratically divergent scalar potential. E.g the mass term in the RHS of \Eq{tildePhiEq} would appear in the original NGB basis as
\begin{align}
\Lambda^2|\lambda_1|^2\text{Tr}[\tilde \Phi \tilde \Phi^*] =& \Lambda^2|\lambda_1|^2(|f_1|^2 \text{Tr}[\Phi\Phi^*]+\frac1f f_1 f_2^* \text{Tr}[H^\dagger \Phi H^*]\nonumber
\\&+\frac1ff^*_1 f_2 \text{Tr}[H^T \Phi^* H]+\frac{1}{f^2}|f_2|^2 (H^\dagger H)^2)\,.
\end{align}
This argument can be repeated for every coupling $c \in \{ \tilde \lambda_1,g_1,g_2,g'_1,g'_2\} $ which generates a scalar potential proportional to $|c|^2 \Lambda^2$. We can immediately see that the symmetry structure allows a quadratically divergent mass term for $\Phi$. The collective breaking structure prevents the appearance of $|H|^2$ in the quadratically divergent potential, as well other operators, such as
\begin{align}
s^2, \varphi^2, s^2 |H|^2\;\; \text{and} \;\; s \vec{H}^\dagger \varphi \vec{H}\,.
\end{align}
Logarithmically divergent 1-loop contributions to the scalar potential contain four couplings.   Possible combinations are trivial combination like $|c|^2 |c'|^2$, and non trivial combinations like
\begin{align}
\lambda_1 \tilde\lambda_1^\dagger \tilde\kappa \kappa^\dagger\,,\,\, \tilde\lambda_1 \lambda_1^\dagger \kappa \tilde\kappa^\dagger + \text{T-conjugates}\,.
\end{align}
At this level all scalar operators can be generated except the singlet mass. The singlet remains exactly massless at 1-loop and must acquire a mass from higher order loops, e.g 2-loop diagram by closing the Higgs loop in the 1-loop induced $s^2 |H|^2$ interaction. We report the radiatively generated couplings calculated from \Eq{fermionCW} after setting all the T-conjugate couplings to their respective values $\tilde c =c$. We neglect the gauge contributions which generate $O(1\%)$ corrections to the fermion loops contribution. We define $C \equiv \frac{N_c}{16 \pi^2}a_2  \log \left(\frac{\Lambda^2 } {f^2 }\right)$. Note that $C \sim 0.1$ for $a_2=1$ and $\Lambda = 4 \pi f$.
\begin{align}
m^2_\Phi &= \frac{ N_c}{4 \pi^2}a_1  |\lambda_1|^2  \Lambda^2\,,
\\
\lambda &=  \frac{ N_c}{16 \pi^2}a_1  |\lambda_1|^2   \left(\frac{\Lambda}{f}\right)^2\,,
\\
 \lambda_{\text{\tiny DM}} &= C \frac{|\lambda_1|^2  r^2 \left(25 |\lambda_2|^2 +6 |\kappa|^2   (r+5)^2\right)}{30 \left(r^2+5\right)} > 2.3 C\,, 
 \\
 \mu^2 &= -C f^2 |\lambda _1|^2 |\lambda _2|^2 < -4C f^2\,,
 \\
 \lambda_{\varphi} &= C \frac{|\lambda_1|^2 r^2 \left(5 |\lambda_2|^2+6 \kappa ^2 (r+1) (r+5)\right)}{3 \sqrt{5} \sqrt{r^4+6 r^2+5}} > 5.2C\,, \label{tripletSingetMixing}
\end{align}
The bounds are calculated assuming $\kappa,r>1$. We find the minimal/maximal value with respect to $\lambda_1,\lambda_2$ under the top Yukawa constraint. The triplet mass is generated only from gauge loops. We define $D\equiv \frac{3}{64 \pi^2}a_4 \log \left(  \frac{\Lambda^2 }{f^2}\right) = \frac{C}{4} \frac{a_4}{a_2}\sim 0.025$. The triplet mass is given by
\begin{align}
m^2_\varphi &= D f^2\left( \frac{8g^4r^2 (1 + r)^2}{1 + r^2}\right) \approx \left(\frac{D}{0.025} \right) \left(\frac{f}{1200~\text{GeV}}\right)^2 \left(850~\text{GeV} \right)^2 \label{tripletMass}
\,, \end{align}
where we used $r=3$. Lastly, we report the Higgs potential. The Higgs potential in the unitary gauge up to order $O(\sin^4 h)$
\begin{align}
V_{h} &=     \frac{N_c}{16 \pi^2}f^2 \Lambda^2a_1 \left( |\lambda _1|^2 +|\tilde\lambda_1|^2 \right) \sin^4\left(\frac{h}{\sqrt{2} f} \right)
\\
&+Cf^4\left( |\tilde\lambda_1|^2|\tilde\kappa|^2  -\tilde\lambda_1 \lambda_1^\dagger\kappa  \tilde\kappa^\dagger -|\tilde\lambda_1|^2|\tilde\lambda_2|^2 +[g \leftrightarrow \tilde g] \right)\sin^2\left(\frac{h}{\sqrt{2} f}
  \right) \nonumber
  \\&+\frac{1}{4}Cf^4 \left(-|\lambda _1|^2| \tilde\lambda_1|^2-|\lambda _1|^2|\lambda _1|^2+2 |\tilde\lambda_1|^2 |\tilde\lambda_2 |^2 \right. \nonumber
  \\&\left.+4 \left(  \lambda_1^\dagger \tilde\lambda_1 \kappa  \tilde\kappa^\dagger - |\tilde\lambda_1|^2 |\tilde\kappa|^2  \right)+[g \leftrightarrow \tilde g]\right)\sin^4\left(\frac{h}{\sqrt{2} f}\right) \,.  \nonumber
\end{align}
After setting $\tilde g = g$, we find
\begin{align}
V_{h} &=   -2Cf^4|\lambda_1|^2|\lambda_2|^2\sin^2\left(\frac{h}{\sqrt{2} f}
  \right)+ \left[ \frac{N_c}{8 \pi^2} \Lambda^2a_1  
 +Cf^2\left(|\lambda_2|^2-|\lambda_1|^2 \right)\right]f^2 |\lambda_1|^2 \sin^4\left(\frac{h}{\sqrt{2} f}\right) \,.\end{align}
Although terms proportional to $|\lambda_1|^2 |\kappa|^2$ could have appeared a priori in the Higgs potential, they vanish due to $T$-parity. Clearly if we were to set $\tilde \kappa  = -\kappa$, which is equivalent to flipping the parity of $\psi^-_R$ and coupling it to the $T$-even combination $\frac{1}{\sqrt{2}}(\psi_1+\psi_2)$, the $\kappa$ coupling would have appeared in the Higgs potential. Since $\kappa$ does not appear in the Higgs potential, taking large values of $\kappa$ would  have no influence on the tuning of the Higgs potential at one loop.
\bibliography{bib/compDmBib}
\bibliographystyle{jhep}

\end{document}